\documentclass[preprint,11pt,authoryear]{elsarticle}
\usepackage[a4paper]{geometry}
\usepackage[singlespacing]{setspace}
\usepackage[round]{natbib} 
\usepackage{amsmath,amsthm,epsfig,amsfonts,color, epstopdf, bm}
\usepackage[ruled,vlined,linesnumbered]{algorithm2e}
\usepackage{amsbsy}
\usepackage{amssymb}
\usepackage{enumerate}
\usepackage{soul}
\usepackage{epstopdf}
\usepackage{rotating}
\usepackage[font=footnotesize,labelfont=bf]{caption}
\usepackage[english]{babel}
\usepackage[utf8x]{inputenc}
\usepackage{graphicx,subfigure} 
\usepackage{booktabs} 
\usepackage{adjustbox}
\usepackage{caption}
\usepackage{makecell}
\usepackage{xcolor}
\usepackage{floatrow}
\usepackage{url}
\usepackage{afterpage}
\usepackage{geometry}
\usepackage{lipsum}
\usepackage{multirow}
\usepackage[flushleft]{threeparttable}
\usepackage{hyperref}
\usepackage{pdflscape}
\usepackage{graphics}
\usepackage{graphicx}
\usepackage{threeparttable}

\newtheorem{definition}{{Definition}}
\newtheorem{proposition}{{Proposition}}
\newtheorem{theorem}{{Theorem}}

\newtheorem{remark}{{Remark}}

\def\sym#1{\ifmmode^{#1}\else\(^{#1}\)\fi}

\newfloatcommand{capbtabbox}{table}[][\FBwidth]

\def\marginnote#1{\setbox0=\vtop{\hsize4pc
\small\raggedright\noindent\baselineskip9pt \rightskip=0.5pc plus
1.5pc #1}\leavevmode \vadjust{\dimen0=\dp0
\kern-\ht0\hbox{\kern-4.00pc\box0}\kern-\dimen0}}
\def\lboxit#1{\vbox{\hrule\hbox{\vrule\kern6pt
\vbox{\kern6pt#1\kern6pt}\kern6pt\vrule}\hrule}}

\usepackage{amssymb}

\journal{arxiv}

\begin{document}

\begin{frontmatter}



\title{Decentralized Quantile Regression for Feature-Distributed Massive Datasets with Privacy Guarantees}


\author[1,2]{Peiwen Xiao}

\author[1]{Xiaohui Liu\corref{cor1}}
\ead{liuxiaohui@jxufe.edu.cn}

\author[2]{Guangming Pan}

\author[3]{Wei Long}

\cortext[cor1]{Correspoding author}

\affiliation[1]{organization={Key Laboratory of Data Science in Finance and Economics, and School of Statistics and Data Science, Jiangxi University of Finance and Economics},
            city={Nanchang}, 
            state={Jiangxi},
            country={China}}

\affiliation[2]{organization={School of Physical and Mathematical Sciences, Nanyang Technological University},,
            country={Singapore}}

\affiliation[3]{organization={Department of Economics, Tulane University}, 
            city={New Orleans},
            state={Louisiana},
            country={United States}}

\begin{abstract}
In this paper, we introduce a novel decentralized surrogate gradient-based algorithm for quantile regression in a feature-distributed setting, where global features are dispersed across multiple machines within a decentralized network. The proposed algorithm, \texttt{DSG-cqr}, utilizes a convolution-type smoothing approach to address the non-smooth nature of the quantile loss function. \texttt{DSG-cqr} is fully decentralized, conjugate-free, easy to implement, and achieves linear convergence up to statistical precision. To ensure privacy, we adopt the Gaussian mechanism to provide $(\epsilon,\delta)$-differential privacy. To overcome the exact residual calculation problem, we estimate residuals using auxiliary variables and develop a confidence interval construction method based on Wald statistics. Theoretical properties are established, and the practical utility of the methods is also demonstrated through extensive simulations and a real-world data application.
\end{abstract}




\end{frontmatter}


\section{Introduction}

In modern data analysis, limitations in data storage capacity, along with spatial isolation and privacy concerns, often make it impractical to process or store entire datasets on a single machine. This scenario presents significant challenges, particularly when raw data is distributed across multiple machines, necessitating the development of novel statistical methodologies within the distributed learning framework. Distributed learning problems can be addressed by assuming the existence of a trusted central machine, employing divide-and-conquer strategies \citep{Xu2017, Chen2019}, or iterative approaches \citep{Tan2021, Fan2023}. These centralized methodologies rely on a trusted central machine that collects and disseminates information to other machines, which, unfortunately, may lead to potential information leakage. Furthermore, within the centralized methodologies, the central machine typically needs to communicate with many machines \citep{Wu2022}, requiring substantial network bandwidth and increasing the risk of network failures. As a solution to these issues, decentralized learning enables communications between neighboring machines within a network, eliminating the need for a central machine. Due to its scalability and robustness, decentralized learning has gained increasing prominence in recent years \citep{Wu2022, Wang2023}.

\begin{figure}[h]
  \centering
  \includegraphics[width=\textwidth]{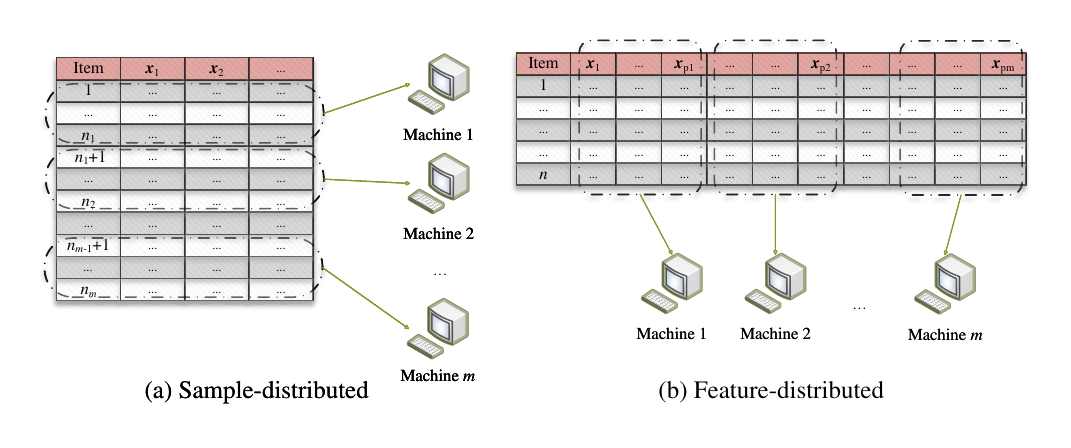}
  \caption{Illustrations of two types of distributed datasets, i.e., sample- and feature-distributed datasets.}
  \label{sdfd}
\end{figure}

Distributed data can be broadly categorized into two types. The first type is sample-distributed data, where each machine holds a subset of the samples. For example, different air quality monitoring sites may collect the same type of air quality data from various locations, with datasets typically not stored on a single machine due to spatial isolation. In this scenario, each machine retains part of the sample information, allowing local computation of gradient information and facilitating the optimization process. Recently, sample-distributed learning has garnered considerable attention and has been extensively studied in the literature, including methods such as gradient tracking \citep{Li2019}, alternating direction method of multipliers (ADMM) \citep{Wang2023}, decentralized gradient descent \citep{Wu2022}, and surrogate likelihood \citep{Tan2021, Jiang2021}. The second type, which is equally important yet less studied in the existing literature, involves feature-distributed data, where the features of a dataset are distributed and stored across different machines, with each machine holding only a subset of the feature information. This scenario is referred to as the feature-distributed case. Feature-distributed datasets are common in real-life applications; for example, a large number of users may log in to various apps (shopping apps, health apps, video apps, etc.) using their personal Google accounts or medical records for the same patients may be stored in different specialized clinics. However, users may prefer not to share their browsing history from certain platforms with other apps linked to the same account, and patients may not want sensitive health information (e.g., confidential diagnoses) to be shared with other clinics. These scenarios share the characteristic that the complete set of features for the same samples is distributed across multiple machines (locations). Additionally, due to privacy and security protocols, raw data from different sources cannot be shared, posing significant challenges for joint modeling. Figure \ref{sdfd} illustrates the difference between the sample- and feature-distributed datasets. 

The growing datasets stored in distributed form under the constraints of information sharing pose challenges to conventional statistical inferential methods, especially in the feature-distributed setting. Recently, \cite{Hu2019} propose an ADMM sharing algorithm to solve the empirical risk minimization problem in the feature-distributed setting, while the proposed ADMM sharing algorithm requires a trusted central node that collects and broadcasts the summary information and thus cannot be extended to the decentralized case, so as the residual projection method studied in \cite{Fan2023}. In the decentralized setting, researchers develop ADMM \citep{Gratton2018} and the decentralized gradient descent \citep{Arablouei2015} algorithms to solve the ordinary least squares problem with ridge penalty over distributed features. Nevertheless, these methods are restricted to regression problems, such as ridge regression, with closed-form solutions, and it seems impossible to generalize them to other cases, e.g., quantile regression, having no closed-form solutions. 

Our research is motivated by the increasing prevalence of distributed features in datasets in the era of big data. Relying exclusively on local information for statistical modeling can lead to model misspecification, particularly when local data is insufficient \citep{Giessing2018}. Moreover, quantile regression (QR), first introduced by \cite{Koenker1978}, is developed to provide a comprehensive understanding of the conditional quantiles of the response variable given certain explanatory variables rather than focusing solely on the mean. By regressing on explanatory variables at different quantiles, QR offers broad insights into the dependent relationships across the conditional distribution of the response variable and is more robust to data with heavy tails than the ordinary least squares method. Formally, the QR model under the feature-distributed setting can be written as 
\begin{eqnarray}\label{qrmodel}
    y_{i} = \sum_{j=1}^m \bm{x}_{ij}^\top \bm \beta_{j, \tau}  + \varepsilon_i, \quad i = 1,\cdots, n; ~j = 1,\cdots, m,
\end{eqnarray}
where $m$ is the number of machines, $n$ is the sample size, $y_{i}$ denotes the scalar response, $\bm{x}_{ij}$ and $\bm \beta_{j, \tau}$ are the $p_j$-dimensional covariates and regression coefficients at the $\tau$-th quantile with $p = \sum_{j=1}^m p_j$. Assuming access to the raw data, the estimator of $\bm{\beta}_{\tau} = (\bm{\beta}_{1,\tau}^\top, \bm{\beta}_{2,\tau}^\top, \cdots, \bm{\beta}_{m,\tau}^\top)^\top$ can be obtained by solving the optimization problem
\begin{eqnarray}\label{qropt}
    \min_{\bm{\beta}_{1, \tau},\cdots, \bm{\beta}_{m, \tau}} \hat{Q}_\tau(\bm{\beta}_\tau) := \min_{\bm{\beta}_{1, \tau},\cdots, \bm{\beta}_{m, \tau}} \frac{1}{n} \sum_{i=1}^n \rho_\tau\left(y_i - \sum_{j=1}^m \bm{x}_{ij}^\top \bm{\beta}_{j, \tau}\right),
\end{eqnarray}
where $\rho_\tau(t) = t(\tau - I_{(t<0)})$ refers to the quantile loss function. It is worth noting that calculation of the exact gradient or residual requires global feature information ($\sum_{j=1}^m \bm{x}_{ij}^\top \bm{\beta}_{j, \tau}$) in the feature-distributed setting, which makes it challenging to conduct optimization, especially in the absence of a trusted central node. One common way to solve this kind of problem is to conduct dual transformation \citep{Alghunaim2020, Gratton2021}, and then optimize the separable objective function. However, the quantile loss function has a non-trivial conjugate function, making these methods difficult to implement in the context of QR. Moreover, existing studies primarily focus on in-sample properties \citep{Hu2019}, leaving population properties for decentralized feature-distributed QR largely unexplored. These challenges underscore the need for decentralized QR methods for feature-distributed datasets, which is the focus of this paper within a decentralized network setting.

Our contributions are threefold. Firstly, we develop a novel decentralized surrogate gradient-based algorithm for QR (\texttt{DSG-cqr}) tailored for feature-distributed datasets. Recently, \cite{Jordan2019} proposed a communication-efficient surrogate likelihood (CSL) framework for sample-distributed datasets. The CSL approach is designed for centralized learning, where a central machine broadcasts the local surrogate loss using global information, making it unsuitable for decentralized optimization in feature-distributed settings. Similarly, the \texttt{Network-DANE} algorithm by \cite{Li2019} operates in a decentralized network but still relies on local gradients to construct surrogate losses, which is infeasible in feature-distributed setting where local gradients cannot be directly computed. In contrast, the proposed \texttt{DSG-cqr} algorithm does not need to calculate the exact local gradient but instead utilizes its surrogate. \texttt{DSG-cqr} also simplifies the optimization process by eliminating the need for nested inner and outer loops, as required by \cite{Jordan2019} and \cite{Li2019}, and is conjugate-free, easy to implement in a decentralized network. To handle the non-smooth nature of the quantile loss function, we employ a convolution-type smoothing technique, proving that \texttt{DSG-cqr} achieves linear convergence to the statistical precision with an appropriate bandwidth selection. We also analyze the detailed impact of algorithmic parameters on its performance. Second, we shift the focus from sample-level error to statistical error analysis within the \texttt{DSG-cqr} framework, advancing statistical inference for QR in feature-distributed datasets. By estimating residuals through auxiliary variables and assuming independent features across machines, we introduce a statistical inference method and establish the theoretical properties of this decentralized inference procedure. This includes the construction of confidence intervals, demonstrating the method’s effectiveness in feature-distributed settings.

Finally, we contribute to the literature by developing a privacy-preserving version of \texttt{DSG-cqr}, and demonstrate that it can effectively balance the trade-off between estimating accuracy and privacy protection. Note that for practitioners, another important concern in distributed statistical modeling is information leakage, as joint modeling typically requires the transmission of at least summary information. A substantial body of literature has highlighted that even summary statistics can lead to the leakage of sensitive information, such as through gradient information \citep{Song2020} or mean information \citep{Kamath2018}. Differential privacy, originally proposed by \cite{Dwork2006}, aims to ensure that the distribution of the random output of a given randomized algorithm remains nearly unchanged when individual data points in the input dataset vary. Differential privacy is a model-free, non-parametric metric that provides a framework to quantify the information leakage of a given procedure or algorithm, which has gained wide attention, particularly in distributed learning \citep{Huang2020, Hu2019}. An algorithm that satisfies $(\epsilon,\delta)$-differential privacy protects the sensitive information of individuals from potential adversaries who may have access to the output. To this end, we adopt the Gaussian mechanism during the iterative process in \texttt{DSG-cqr}, and the resulting algorithm satisfies $(\epsilon,\delta)$-differential privacy property. We establish an upper bound for the estimation error, accounting for the distribution of samples and the added Gaussian noise for the feature-distributed dataset in a decentralized network. We also derive the ``optimal privacy combination'' for $(\epsilon, \delta)$, providing the best privacy protection with the constraint of maintaining statistical accuracy.

The rest of this paper is organized as follows. Section \ref{sec:method} provides detailed methodologies and main results. Simulations and empirical data analysis are conducted in Section \ref{sec:sim} and Section \ref{sec:emp}, respectively. We conclude in Section \ref{sec:conclu}. To save space, detailed technical proofs of the main results, and additional simulation results are included in the appendix. 

\section{Methodology and Main Results}
\label{sec:method}

\subsection{Notations and Preliminary}

We start with some notations and preliminaries that will be used throughout this paper. In the sequel, the Frobenius norm is denoted by $\Vert \cdot \Vert_F$, and $\Vert \cdot \Vert_2$ denotes the largest eigenvalue of a matrix. The $l_1$ and $l_2$ norms of a vector are represented by $\Vert \cdot \Vert_1$ and $\Vert \cdot \Vert_2$, respectively. The symbol $\otimes$ refers to the Kronecker product, and $\langle \cdot, \cdot \rangle$ represents the inner product of two vectors of the same length. Functions $\lfloor\cdot\rfloor$ and $\lceil\cdot\rceil$ denote the floor and ceiling functions, respectively. $I_{(\cdot)}$ denotes the indicator function, and $\bm{1}_p$ represents the $p$-dimensional all-ones vector. The determinant of a square matrix is denoted by $\mathrm{det} (\cdot)$. Term $\log$ refers to the natural logarithm, while $\log_\alpha$ represents the logarithm with base $\alpha$. For any two positive sequences $\{a_n\}$ and $\{b_n\}$, $a_n \lesssim b_n$ or $a_n = O(b_n)$ implies that $a_n \leq Cb_n$ for some positive constant $C$ when $n$ is sufficiently large. `$\stackrel{d}\to$' and `$\stackrel{p}\to$' stand for the convergence in distribution and in probability, respectively.

Furthermore, we define a network with graph $\mathcal{G} = (\mathcal{V}, \mathcal{E})$, where $\mathcal{V} = \{1,2,\cdots, m\}$ is the set of nodes/machines, and $\mathcal{E} \subseteq \mathcal{V}\times \mathcal{V}$ is the set of edges. We assume that $\mathcal{G}$ is undirected and connected, which means $(i, j) \in \mathcal{E}$ can induce $(j, i) \in \mathcal{E}$, and there is a path between any two nodes $i, i' \in \mathcal{V}$. In the decentralized optimization framework under the feature-distributed setting, the response variable is accessible to all machines. Data features (predictors) are distributed over $m$ nodes/machines. Each node $i\in \mathcal{V}$ can only communicate with its neighbors $\mathcal{N}_i$, where $\mathcal{N}_i = \{i': w_{ii'} \neq 0\}$ as the neighborhood set of the $i$-th machine.

\textit{Dynamic Consensus Theory}. For convenience, we introduce a $m\times m$ mixing matrix $\bm{W} = [w_{ii'}]_{1\leq i, i' \leq m}$, where each $(i, i') \in \mathcal{E}$ is associated with a non-negative weight $w_{ii'}$. Specifically, $w_{ii'} = 0$ indicates that machines $i$ and $i'$ are not connected. Figure \ref{network} shows three examples of network structures. For clarity, we impose the following synchronization assumption, C1), on the network and mixing matrix $\bm{W}$, which is a commonly used condition in decentralized distributed optimization \citep{Arablouei2015, Choi2022}:

\begin{figure}[h]
  \centering
  \includegraphics[width=\textwidth]{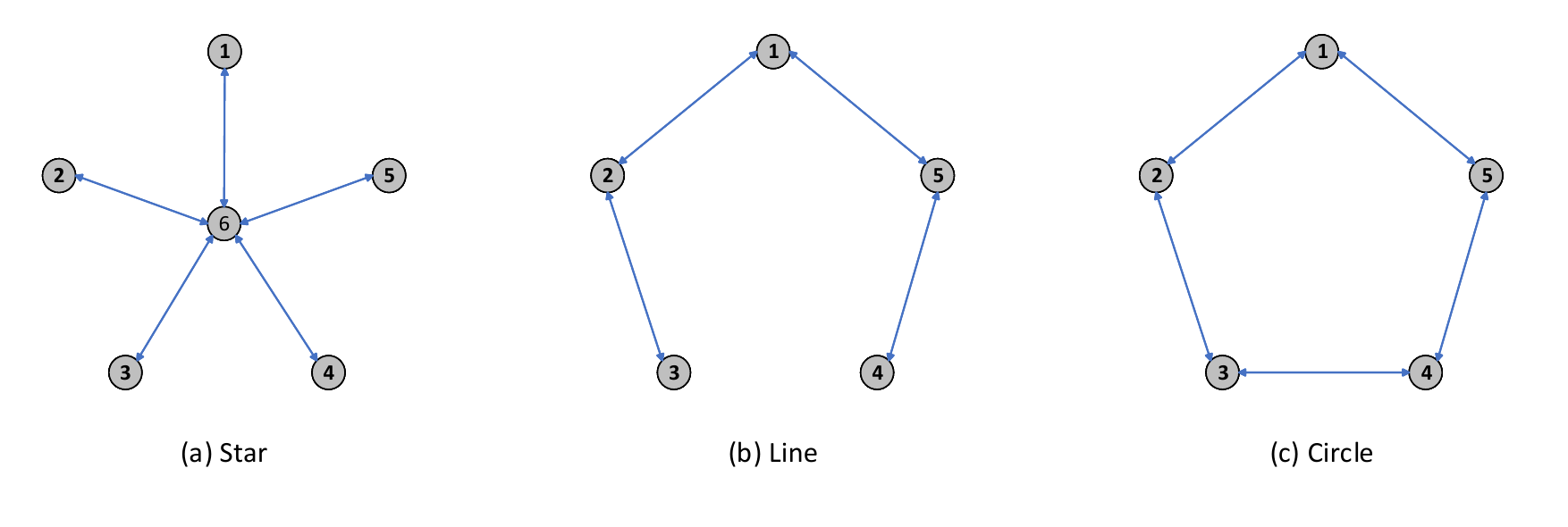}
  \caption{Examples of common network topology structures. The left, middle, and right panels refer to the Star, Line, and Circle networks, respectively.}
  \label{network}
\end{figure}

\noindent C1) Assume the network has a synchronized clock. The mixing matrix $\bm{W}$ is doubly stochastic, that is, $\sum_i w_{ij} = \sum_j w_{ij} = 1$. Denote 
$$\alpha = \left\Vert \bm{W} - \frac{1}{m}\bm{1}_m\bm{1}_m^\top \right\Vert_2.$$

By the structure of $\bm{W}$, it is easy to check that the largest eigenvalue of $\bm{W}$ is 1, with its corresponding eigenvector being $\bm{1}_m$ and the second largest eigenvalue of $\bm{W}$ is $\alpha$, which lies within the interval $[0,1]$. Note that $\alpha$ can be interpreted as a measure of network sparsity, with larger values of $\alpha$ indicating a sparser network. The synchronized clock implies that the parameter updates for each machine occur simultaneously within each iteration. Under condition C1) and the spectral decomposition theorem, we have $\lim_{\kappa_0\to\infty}\bm{W}^{\kappa_0} = \frac{1}{m}\bm{1}_m\bm{1}_m^\top$. Consequently, if machine $j \in \mathcal{V}$ holds a $n$-dimensional vector $\bm{q}_j \in \mathbb{R}^n$, after a sufficiently large number of communication rounds with its neighbors, based on the mixing matrix $\bm{W}$—where each machine updates its value by taking a weighted average of the values of its neighbors—the value of each machine will converge to the global mean \citep{OlfatiSaber2007}:
$$\lim_{\kappa_0\to\infty}\bm{q}\bm{W}^{\kappa_0} = \bm{q}\frac{1}{m}\bm{1}_m\bm{1}_m^\top = \frac{1}{m}\sum_{j=1}^m\bm{q}_j \bm{1}_m^\top,$$
where $\bm{q} = ( \bm{q}_1, \bm{q}_2, \cdots, \bm{q}_m )$. This is the so-called Dynamic Consensus Theory, which has been widely applied in decentralized sample-distributed optimization \citep{Wu2022, Choi2022}. In the sequel, we will explore its potential application to the feature-distributed setting in the methodology section.

\subsection{Decentralized Quantile Regression for Feature-Distributed Data}

In this subsection, we discuss the computational algorithm and convergence rate of the resulting estimator of the decentralized quantile regression for feature-distributed data. For ease of exposition, we omit the subscript $\tau$ in the regression coefficients.

Note that the estimation of true coefficients $\bm{\beta}_0$ (under quantile level $\tau$) for model \eqref{qrmodel} usually depends on computing the sub-gradient of the quantile loss function, i.e., $\nabla \hat{Q}_\tau(\bm{\beta}) = -\frac{1}{n} \sum_{i=1}^n \psi_\tau(y_i - \sum_{j=1}^m \bm{x}_{ij}^\top \bm{\beta}_j)\bm{x}_i \in \mathbb{R}^p$, where  $\psi_\tau(t) = \tau - I_{(t<0)}$ is a two-valued function. The computation is possible under the non-distributed setting as all datasets are stored on a single machine and can be read or written into memory. However, since $\nabla \hat{Q}_\tau(\bm{\beta})$ contains the quantity $\sum_{j=1}^m \bm{x}_{ij}^\top \bm{\beta}_j$, which requires data aggregation, conventional vanilla (sub-)gradient descent algorithms that rely on $\nabla \hat{Q}_\tau(\bm{\beta})$ are inapplicable in the feature-distributed setting. When local datasets $\{\{\bm{x}_{ij}\}_{j=1}^m\}_{i=1}^n$ are stored across different machines, shared memory access is not possible, making it challenging to optimize the quantile loss function in a decentralized network.

To address this issue, we introduce the concept of a surrogate sub-gradient. Specifically, we replace the exact local sub-gradient $-\frac{1}{n} \sum_{i=1}^n\psi_\tau(y_i - \sum_{j=1}^m \bm{x}_{ij}^\top \bm{\beta}_j)\bm{x}_{ij}$ with a surrogate sub-gradient $-\frac{1}{n} \sum_{i=1}^n\psi_\tau(y_i - mz_{ij})\bm{x}_{ij}$ within each iteration, where $\bm{z}_{j} = (z_{1j}, z_{2j},\cdots, z_{nj})^\top$ denotes an auxiliary variable assumed to be computable for the $j$-th machine. The motivation for doing so is as follows: As long as the auxiliary variables $\{z_{ij}\}_{j=1}^m$ closely approximate the product $\frac{1}{m}\sum_{j=1}^m \bm{x}_{ij}^\top \bm{\beta}_j$, the surrogate sub-gradient descent algorithm can perform comparably to the centralized approach. However, simply replacing $\{\bm{x}_{ij}^\top \bm{\beta}_j\}$ by $\{z_{ij}\}$ is insufficient, as analyzing the difference $\Vert\psi_\tau(y_i - \sum_{j=1}^m \bm{x}_{ij}^\top \bm{\beta}_j) - \psi_\tau(y_i - mz_{ij})\Vert_2$ proves challenging due to the non-smooth nature of the quantile loss function. To address this, we adopt a \underline{con}volution smoothed \underline{qu}antile \underline{re}gression, or \texttt{conquer}, as suggested by \cite{Fernandes2021}, to effectively handle this non-smoothness issue. Under \texttt{conquer}, we estimate $\bm{\beta}_0$ through minimizing
\begin{equation}\label{smqr_opt}
    \min_{\bm{\beta}_1,\cdots, \bm{\beta}_m}\hat{Q}_{\tau,h}(\bm{\beta}) = \min_{\bm{\beta}_1,\cdots, \bm{\beta}_m} \frac{1}{n} \sum_{i=1}^n \rho_{\tau,h}\left(y_i - \sum_{j=1}^m \bm{x}_{ij}^\top \bm{\beta}_j\right),
\end{equation}
$$\text{with}~~\rho_{\tau,h} = (\rho_\tau * K_h)(u) := \int_{-\infty}^{\infty}\rho_\tau(v) K_h(v-u)dv,$$
where $K(\cdot)$ is a non-negative kernel function, $*$ denotes the convolution operator, and $K_h(u) = \frac{1}{h} K(u/h)$ with $h > 0$ being a selected bandwidth which may shrink to zero as the sample size grows to ensure the statistical accuracy and computational stability. With a non-negative kernel function, the empirical smoothed quantile loss function $\hat{Q}_{\tau,h}(\bm{\beta})$ is twice continuously differentiable with the gradient and Hessian matrix defined as
\begin{eqnarray}\label{grad}
    \nabla \hat{Q}_{\tau, h}(\bm{\beta}) := \begin{pmatrix}
        \nabla \hat{Q}_{\tau, h}(\bm{\beta}_1) \\ \nabla \hat{Q}_{\tau, h}(\bm{\beta}_2) \\ \vdots \\ \nabla \hat{Q}_{\tau, h}(\bm{\beta}_m)
    \end{pmatrix} = \frac{1}{n} \sum_{i=1}^n \left\{\Bar{K} \left(\left(\sum_{j=1}^m \bm{x}_{ij}^\top \bm{\beta}_j - y_i\right) / h\right) - \tau \right\}\begin{pmatrix}
        \bm{x}_{i1} \\ \bm{x}_{i2} \\ \vdots \\\bm{x}_{im}
    \end{pmatrix} \in \mathbb{R}^p,
\end{eqnarray}

$$\nabla^2 \hat{Q}_{\tau, h}(\bm{\beta}) = \frac{1}{n}\sum_{i=1}^n K_h\left(y_i - \sum_{j=1}^m \bm{x}_{ij}^\top \bm{\beta}_j\right) \bm{x}_i\bm{x}_i^\top,$$
where $\Bar{K}(u) = \int_{-\infty}^u K(v)dv$, $\bm{x}_i = \left(\bm{x}_{i1}^\top, \bm{x}_{i2}^\top, \cdots, \bm{x}_{im}^\top\right)^\top$. Hence, by replacing the incalculable product $\frac{1}{m}\sum_{j=1}^m \bm{x}_{ij}^\top \bm{\beta}_j$, we may further define the smoothed surrogate gradient
\begin{eqnarray}\label{sur_grad}
    \nabla \hat{S}_{\tau, h}(\bm{z}^{(t)}) := \begin{pmatrix}
        \nabla \hat{S}_{\tau, h}(\bm{z}^{(t)}_1) \\ \nabla \hat{S}_{\tau, h}(\bm{z}^{(t)}_2) \\ \vdots \\ \nabla \hat{S}_{\tau, h}(\bm{z}^{(t)}_m)
    \end{pmatrix} = \begin{pmatrix}
        \frac{1}{n} \sum_{i=1}^n \left\{\Bar{K} \left((m\bm{z}_{i1}^{(t)} - y_i) / h\right) - \tau \right\}\bm{x}_{i1} \\ \frac{1}{n} \sum_{i=1}^n \left\{\Bar{K} \left((m\bm{z}_{i2}^{(t)} - y_i) / h\right) - \tau \right\}\bm{x}_{i2} \\ \vdots \\ \frac{1}{n} \sum_{i=1}^n \left\{\Bar{K} \left((m\bm{z}_{im}^{(t)} - y_i) / h\right) - \tau \right\}\bm{x}_{im}
\end{pmatrix}  \in \mathbb{R}^p,
\end{eqnarray}
which will be frequently used in the rest of this paper.

\begin{remark}\label{rmk_smqr}
In the literature, alternative smoothing methods beyond \texttt{conquer} can also be used to construct smoothed surrogate gradients. For example, one straightforward approach is to apply quadratic smoothing within a $h$-neighborhood of 0 and consider the asymmetric Huber loss \citep{Ichinose2023}. However, the Huber-type quantile loss function is only locally strongly convex within this $h$-neighborhood, posing challenges for optimization and solution stability, particularly when $h$ is sufficiently small. Additionally, \cite{Horowitz1996} proposes a smoothed quantile loss function defined as $\rho_\tau^{Horo}(u) = u\left(\tau - \Bar{K}(-u/h)\right)$, where $\Bar{K}$ is defined as previously described. However, Horowitz's smoothed quantile loss function is non-convex, making it extremely difficult to optimize in high-dimensional settings. In contrast, the \texttt{conquer} loss is strictly convex when a non-negative kernel is employed, making it a more suitable choice.
\end{remark}

Based on the local auxiliary variables and the \texttt{conquer} smoothing technique, we propose to optimize \eqref{smqr_opt} via an iterative procedure, whose pseudo-code is provided in Algorithm \ref{alg:algorithm1} below; we refer to this method as \texttt{DSG-cqr}. The motivation for using such an iterative procedure is that the global convergence of the optimization problem in \eqref{smqr_opt} via the vanilla gradient descent method is not significantly impacted by negligible errors or gradually vanishing biases within each update step. Accordingly, we update the local parameters for each machine synchronously using the surrogate gradient, induced by the local auxiliary variable $\{\bm{z}_{j}\}_{j=1}^m$. Additional tracking and mixing steps with a finite number of $\kappa_0$ rounds of the auxiliary variables are performed to ensure that the surrogate bias of the gradient diminishes iteratively. These two parts form the core of Algorithm \ref{alg:algorithm1}.

\begin{algorithm}[h]\footnotesize
	\caption{\footnotesize Decentralized \texttt{conquer} for feature-distributed datasets.}
	\label{alg:algorithm1}
	\KwIn{Data $(\bm{x}_{ij},y_{i})_{i=1,\cdots,n~j=1,\cdots,m}$, quantile level $\tau$, step size $\eta$, bandwidth $h$, mixing matrix $\bm{W} = [w_{ii'}]_{1\leq i, i' \leq m}$, mixing rounds $\kappa_0$, and maximum step $T_{\max}$.}
    
	\BlankLine
    Set step $t = 0$, initial estimates $\bm{\beta}_j^{(0)}$, and initial auxiliary variables $\bm{z}_j^{(0)} = \bm{x}_j^\top\bm{\beta}_j^{(0)}$ for the $j$-th machine, $ 1\leq j \leq m$. 
    
    \While{not converge \text{or} $t < T_{\max}$}{
    Update the step: $t = t+1$.
        
        \For{machine j in $1,2,\cdots,m$}{
        Update the parameter estimate by:
        $$\bm{\beta}_j^{(t)} = \bm{\beta}_j^{(t-1)} - \eta \nabla \hat{S}_{\tau, h}\left(\bm{z}^{(t-1)}_j\right).$$
        
        Set $\bm{z}_j^{(t), 0} = \bm{z}_j^{(t-1)} - \eta \bm{x}_j^\top\nabla \hat{S}_{\tau, h}\left(\bm{z}^{(t-1)}_j\right).$
        }
        \For{machine j in $1,2,\cdots,m$}{
            \For{k in $1,2,\cdots, \kappa_0$}{
            Receive auxiliary variables from the neighborhoods and aggregate:
            $$\bm{z}_j^{(t), k} = \sum_{i\in \mathcal{N}_j}w_{ij}\bm{z}_i^{(t-1), k-1} .$$
	      }
        Set the updated $\kappa_0$-mixing auxiliary variable $\bm{z}_j^{(t)} = \bm{z}_j^{(t), \kappa_0}$.
	  }
	}
    \KwOut{Estimator $\left\{\bm{\beta}^{(T)}_j\right\}_{j=1}^m$, where $T$ denotes the final iterative number.}  
\end{algorithm}

The main steps of \texttt{DSG-cqr} are detailed as follows. Initially, each $j$-th machine starts with its own initialized estimate $\bm{\beta}_j^{(0)}$ and auxiliary variable $\bm{z}_j^{(0)}$. The initial estimates may be chosen based on prior knowledge, such as isolated estimators derived from local data. The tracking step (step 6) and the initial auxiliary variable satisfying $\bm{z}_j^{(0)} = \bm{x}_j^\top\bm{\beta}_j^{(0)}$ together ensure the equation $\sum_{j=1}^m \bm{z}_j^{(t)} = \sum_{j=1}^m \bm{x}_j^\top\bm{\beta}_j^{(t)}$ holds at each iteration. During steps 7-10, each machine broadcasts its auxiliary variable and receives auxiliary variables from its neighborhood machines, then updates its auxiliary variable using a preset doubly stochastic matrix $\bm{W}$. Through iterative execution of these steps, all auxiliary variables converge towards $\lim\limits_{t\to\infty}\frac{1}{m}\sum_{j=1}^m \bm{x}_j^\top\bm{\beta}_j^{(t)}$. Consequently, the surrogate gradient, driven by these auxiliary variables, progressively aligns with the steepest descent direction of the global loss function, leading to convergence. Throughout the iteration of \texttt{DSG-cqr}, communication occurs exclusively in steps 7-10 between neighboring machines, involving the transmission of auxiliary variables rather than raw data, thereby enhancing the protection of local sensitive information. For more details and extensions of Algorithm \ref{alg:algorithm1}, we refer readers to Appendix.

Next, we investigate the theoretical properties of the resulting estimator obtained by Algorithm \ref{alg:algorithm1}. The following conditions are needed.
\begin{enumerate}
\item[C2)] Local predictors $\bm{x}_j \in \mathbb{R}^{p_j}, 1\leq j \leq m$, are sub-Gaussian with sub-Gaussian norm $v_j$. $\bm{\Sigma} = \mathbb{E}(\bm{x}\bm{x}^\top)$ is a positive definite matrix with all eigenvalues in $[\sigma_l, \sigma_u]$. 

\item[C3)] $K(u)$ is a symmetric and non-negative bounded kernel function that integrates to 1, that is, $K(u) = K(-u), K(u) \geq 0~\forall u \in \mathbb{R}, \int_{-\infty}^{\infty} K(u) du = 1$, and $\int u^2 K(u) du < \infty$. 

\item[C4)] There exists positive constants $\underline{f}, \overline{f}, l_f$, such that $\underline{f} \leq f_{\varepsilon|\bm{x}}(\cdot) \leq \overline{f}$ and $|f_{\varepsilon|\bm{x}}(a) - f_{\varepsilon|\bm{x}}(b)| \leq l_f |a-b|$ for all $a, b \in \mathbb{R}$ almost surely (over $\bm{x}$).

\item[C5)] There exist two constants $0<a_l<a_u<\infty$, such that all the eigenvalues of the population Hessian matrix $\mathbf{\bm{H}}(\bm{\beta}) = \mathbb{E}\left\{f_{\varepsilon|\bm{x}}(y-\bm{x}^\top\bm\beta) \bm{x}\bm{x}^\top\right\}$ lie in $[a_l, a_u]$ almost surely (over $\bm{x}$).
\end{enumerate}
The sub-Gaussian condition in C2) is imposed to guarantee exponential-type concentration probabilistic bounds for the norm of the covariates \citep{Tan2021}. C3) and C4) are common assumptions for the kernel function and conditional density assumptions for the quantile regression model, respectively; both are necessary to characterize the smoothing error explicitly \citep{He2023}. C5) imposes strong convexity on the population loss, a typical requirement in decentralized gradient descent algorithms \citep{Li2019, Wu2022}.

The following theorem states the convergence result of $\bm{\beta}^{(T)} = \left(\left(\bm{\beta}^{(T)}_1\right)^\top, \cdots, \left(\bm{\beta}^{(T)}_m\right)^\top\right)^\top$ obtained by using the \texttt{DSG-cqr} algorithm.

\begin{theorem}\label{th1}
Assume that conditions C1) - C5) hold, let $\bm{\beta}^{(0)}$ be an initial estimate of order $O(1)$, and if the step size satisfies $0 < \eta \leq \overline{\eta}$, with $\overline{\eta} := \min\left\{ \frac{1-\alpha^{\kappa_0}}{\alpha^{\kappa_0}} \frac{a_{ul}}{\overline{f}m\sigma_u} ,\frac{a_{ul}}{a_l},\frac{1}{a_u} \right\}$, where $a_{ul} = \left(\frac{a_u}{a_l + a_u}\right)^{1/2} \in (0, 1)$. Then, after a fixed number of $T$ iterations, we have 
\begin{equation*}
    \Vert \bm{\beta}^{(T)} - \bm{\beta}_0 \Vert_2 \lesssim \rho^T + \frac{\eta (r^*_1 + r^*_2)}{1-\rho} ~~ \mathrm{and} ~~ \Vert \bm{\beta}^{(T)}_j - \bm{\beta}_{0, j} \Vert_2 \lesssim \rho^T + \frac{\eta (r^*_2 + r^*_{3, j})}{1-\rho}
\end{equation*}
with probability at least $1-CTn^{-1}$, where $C$ is a positive constant independent of $n,p,m$. The decay factor $\rho = \max\left\{ 1 - \eta a_{ul} a_l,~ \alpha^{\kappa_0}(1 + \eta a_{ul} \overline{f}m\sigma_u ) \right\}$ lies between 0 and 1, and $r^*_1$, $r^*_2$ are the remaining terms with $r^*_1 \asymp \sqrt{\frac{p+\log n}{n}}, r^*_2 \asymp h^{2}, r^*_{3, j} \asymp \sqrt{\frac{p_j+\log n}{n}}$, respectively.
\end{theorem}

The detailed proof of Theorem \ref{th1} is relegated to Appendix. This theorem demonstrates that our algorithm can achieve linear convergence up to the level of statistical precision, both globally and locally, with the appropriate selection of bandwidth $h$. Regarding the decay factor $\rho$, due to insufficient mixing, we observe that $\rho$ is always greater than the corresponding value in the centralized case, where the decay factor is $1-\eta a_l$ \citep{Boyd2004b}. Additionally, the step size $\eta$ satisfies $0 < \eta \leq \overline{\eta}$, where $\overline{\eta} = O(\frac{1}{m})$. If the number of machines is held constant, the influence of $m$ on the remaining terms can be considered negligible. Moreover, in scenarios where $m$ diverges, $1-\rho$ and $\eta$ are both of order $O(\frac{1}{m})$, meaning their combined effect remains negligible. Thus, our algorithm exhibits insensitivity to the number of machines.

The decay factor $\rho$ is also influenced by the number of mixing rounds ${\kappa_0}$. Specifically, increasing ${\kappa_0}$ enhances the consensus step, resulting in smaller surrogate bias and faster convergence, albeit at the cost of requiring more communication rounds within each iteration. This indicates a potential trade-off between consensus accuracy and the number of communication rounds. Consequently, it might be possible to reduce communication costs while maintaining statistical efficiency by appropriately selecting ${\kappa_0}$, particularly in sparse network settings. The following theorem establishes the statistical efficiency of the proposed algorithm and determines the optimal ${\kappa_0}$ that minimizes communication costs to achieve it.

\begin{theorem}\label{th2}
Assume that the conditions of Theorem \ref{th1} hold, and let $h \asymp \left( \frac{q+\log n}{n} \right)^{\gamma}$ for any $\gamma \in [1/3, 1/2]$. If $p_j = O(q),~ \forall j$, then after $T \asymp \left\lceil \log\left(\frac{n}{q+\log n}\right)\right\rceil$ steps of iterations, the \texttt{DSG-cqr} estimator $\Tilde{\bm{\beta}} := \bm{\beta}^{(T)}$ satisfies
\begin{equation}\label{th2eq1}
    \Vert \Tilde{\bm{\beta}} - \bm{\beta}_0 \Vert_2 \lesssim \sqrt{\frac{p+\log n}{n}} ~\mathrm{\textit{and}} ~ \Vert \Tilde{\bm{\beta}}_j - \bm{\beta}_{0, j} \Vert_2 \lesssim \sqrt{\frac{q+\log n}{n}},
\end{equation}
with probability at least $1-CTn^{-1} \to 1$, as $n \to \infty$ and $p + \log n \lesssim n$. Furthermore, by choosing $\kappa_{0,\mathrm{opt}} = \left\lfloor\log_\alpha \left(\frac{1-\eta a_{ul} a_l}{1 + \eta a_{ul} \overline{f}m\sigma_u} \right)\right\rfloor \vee 1,$ it takes the minimum communication rounds to achieve \eqref{th2eq1}. 
\end{theorem}

Theorem \ref{th2} shows that $\Tilde{\bm{\beta}}$ can achieve the same statistical efficiency as the centralized counterpart \citep{He2023}, after a logarithmic number of iterations, given an appropriately selected bandwidth $h$. Additionally, it is important to note that the isolated quantile estimator achieves an estimation error of  $O\left(\sqrt{\frac{q + q\log m+\log n}{n}}\right)$ as $n \to \infty$ \citep{Giessing2018}, which is sensitive to the number of machines due to potential model misspecification. In contrast, our solution provides a local estimator with an estimation error of order $O\left(\sqrt{\frac{q+\log n}{n}}\right)$ as $n \to \infty$, thereby improves the estimation error by leveraging local information.

\begin{figure}[h]
  \centering
  \includegraphics[width=0.8\textwidth]{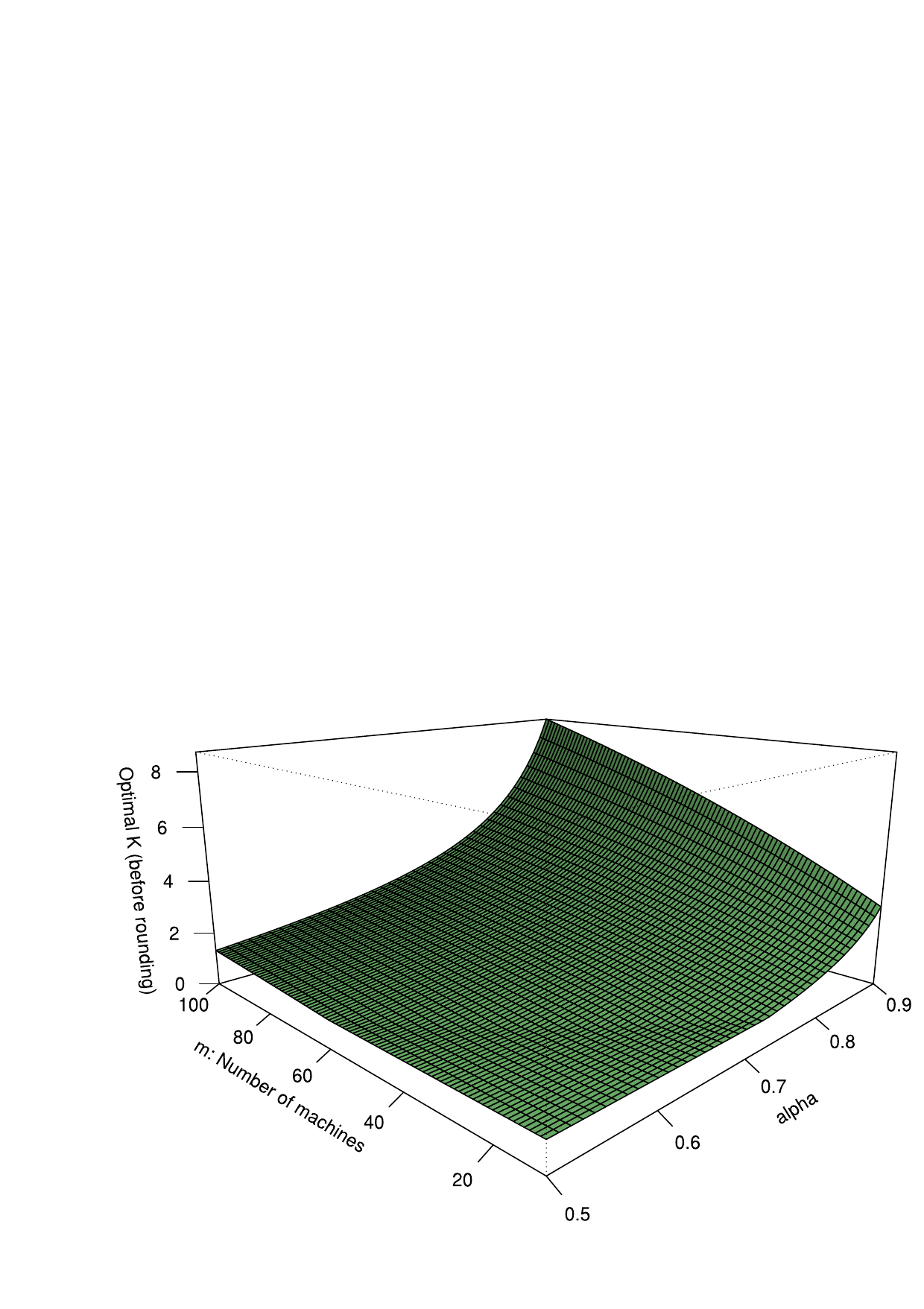}
  \caption{The surface plot of $h(\alpha, m) := \log_\alpha\left(\frac{0.8}{1 + 0.01m}\right) \vee 1$, where $m$ denotes the number of machines.}
  \label{Kopt}
\end{figure}

Theorem \ref{th2} suggests that, apart from the population-related parameters, a larger number of machines and a sparser network necessitate more rounds of mixing to achieve statistical efficiency with a minimal number of communication rounds. Figure \ref{Kopt} presents a three-dimensional surface plot illustrating the optimal number of mixing rounds, $\kappa_{0,\mathrm{opt}}$, as a function of the network structure parameters $\alpha$ and $m$, without flooring. The figure indicates that, in most cases, relatively few mixing rounds are sufficient to achieve fast convergence rather than requiring a significantly large ${\kappa_0}$. Similar findings can also be observed in the subsequent simulation in Section~\ref{sec:sim}.

\subsection{Differential Privacy}

Although \texttt{DSG-cqr} enables effective joint modeling under a feature-distributed setting without raw data communication, statistics (in the form of auxiliary variables) must still be transmitted between neighboring machines during each update step. To protect sensitive information in distributed learning, the local differential privacy \citep{Kairouz2014} metric is widely considered. Unlike traditional differential privacy, which directly perturbs the output, local differential privacy aims to protect locally sensitive information: Each local machine runs a random perturbation before interacting with other machines. Formally, the local differential privacy is defined as follows:

\begin{definition}\label{edp}
(Local Differential Privacy, LDP) A randomized algorithm $\mathcal{M}$ satisfies $(\epsilon,\delta)$-local differential privacy if for any two neighboring datasets, denoted as $\mathcal{D}_j$ and $\mathcal{D}'_j$ and owned by the $j$-th machine $\in \mathcal{V}$, satisfy $\mathbb{P}_d(\mathcal{M}(\mathcal{D}_j) \in \mathcal{A}) \leq e^\epsilon \mathbb{P}_d(\mathcal{M}(\mathcal{D'}_j) \in \mathcal{A}) + \delta,$ where $\mathbb{P}_d$ represents the probability calculation taken conditional on the dataset, $\mathcal{A}$ is an arbitrary subset of $\mathrm{range}(\mathcal{M})$, and the neighboring datasets $\mathcal{D}_j, \mathcal{D}'_j$ differ in only one sample. 
\end{definition} 

Intuitively, for a non-randomized algorithm, it is possible to infer individual raw data from the output statistics in conjunction with other raw data information. Therefore, incorporating randomness is essential to protect individual sensitive information \citep{Dwork2006}. Under the synchronization condition C1), the transmitted auxiliary variable in the \texttt{DSG-cqr} algorithm is a function of the input datasets. To this end, we add independent Gaussian noises to the local surrogate gradient in each step, that is, replace the updating steps 5 and 9 in Algorithm \ref{alg:algorithm1} to:
\begin{eqnarray}\label{updr_prvcy}
        \left\{
    \begin{array}{lcl}
            \bm{\beta}_{\mathrm{private},j}^{(t)} = \bm{\beta}_{\mathrm{private},j}^{(t-1)} - \eta \left(\nabla \hat{S}_{\tau, h}\left(\bm{z}^{(t-1)}_{\mathrm{private},j}\right) + \delta_j^{(t)}\right),   \\
            \bm{z}_{\mathrm{private},j}^{(t), 0} = \bm{z}_{\mathrm{private},j}^{(t-1)} - \eta\bm{x}_{j}^\top\left(\nabla \hat{S}_{\tau,j}\left(\bm{z}^{(t-1)}_{\mathrm{private},j}\right) + \delta_j^{(t)}\right),
    \end{array}\right.
\end{eqnarray}
where $\delta_j^{(t)}$ is a zero mean random noise generated from a well-designed Gaussian distribution. The following theorem ensures that the auxiliary variables $\bm{z}_{\mathrm{private},j}^{(t)}$ in the above process satisfy $(\epsilon, \delta)$-local differential privacy during transmission under the decentralized network.

\begin{theorem}\label{thdp}
Let $\bm{x}_{\cdot j}$ be the $n\times p_j$ sub-sample of the $j$-th machine. Assume that the conditions C1) - C5) hold, and for any two neighborhood datasets (samples) $\bm{x}_{\cdot j}$, $\bm{x}_{\cdot j}'$, i.e., there exists $\Delta_{j} = \sup\limits_t\sup\limits_{\bm{x}_{\cdot j}, \bm{x}_{\cdot j}'} \Vert \bm{z}_{j, \bm{x}_{\cdot j}}^{(t),0} - \bm{z}_{j, \bm{x}_{\cdot j}'}^{(t),0} \Vert_2$, with $\bm{z}_{j, \bm{x}_{\cdot j}}^{(t), 0}$ being the calculation based on $\bm{x}_{\cdot j}$. Then, if the added noises are i.i.d. generated from $N(\bm{0}, 2\epsilon^{-2}\Delta_{j}^2\log(1.25/\delta)(\bm{x}_{\cdot j}\bm{x}_{\cdot j}^\top)^{-1})$,
with an arbitrary constant $\epsilon \in (0,1]$, \texttt{DSG-cqr} algorithm guarantees $(\epsilon, \delta)$-local differential privacy within each iteration. Specifically, 
$$\mathbb{P}_d\left(\bm{z}_{\mathrm{private},j}^{(t),k} | \bm{x}_{\cdot j}\right) \leq e^{\epsilon}\mathbb{P}_d\left(\bm{z}_{\mathrm{private},{\cdot j}}^{(t),k} | \bm{x}_{\cdot j}'\right) + \delta, ~~ \forall 1\leq j \leq m,~~ 0 \leq k \leq {\kappa_0}-1,$$
for $t = 1, 2,\cdots$, where $\mathbb{P}_d$ refers to the probability calculations taken on the added noises.
\end{theorem}
Theorem \ref{thdp} demonstrates that $(\epsilon, \delta)$-differential privacy is maintained at each iteration by adding appropriately designed Gaussian noise.

\begin{remark}\label{rmk_pv}
    The $l_2$ sensitivity term $\Delta_{j}$ plays an important role in the noise calibration process, though calculating its explicit form is highly challenging. Moreover, the $l_2$ sensitivity decreases exponentially as the auxiliary variables reach consensus throughout the iteration process. Therefore, let $\hat{c}_{\bm{x}_j} = \max\limits_{i} \Vert \bm{x}_{ij} \Vert_2$, we recommend using a time-varying, empirical estimate $\hat{\Delta}_{j}^{(t)}$ during the iteration of \texttt{DSG-cqr} in practice, a method that will be adopted in the simulation section:
    \begin{eqnarray}\label{hp_par}
       \hat{\Delta}_{j}^{(t)} = \left\{
    \begin{array}{lcl}
            2\hat{c}_{\bm{x}_j} \left \Vert \bm{\beta}_j^{(0)}\right\Vert_2, & \mathrm{if} ~ t = 1, \\
            2\hat{c}_{\bm{x}_j} \left \Vert \nabla\hat{S}_{\tau,j}(\bm{z}^{(t-1)}_j) \right\Vert_2, & \mathrm{if} ~t > 1.
    \end{array}\right.
    \end{eqnarray}
\end{remark}

\begin{theorem}\label{codp}
    Assume that the conditions of Theorems \ref{th2} and \ref{thdp} hold. Then, after $T \asymp \left\lceil \log\left(\frac{n}{q+\log n}\right)\right\rceil$ steps of update \eqref{updr_prvcy}, the local private estimator of the $j$-th machine $\Tilde{\bm{\beta}}_{\mathrm{private},j} := \bm{\beta}_{\mathrm{private},j}^{(T)}$ satisfies 
    \begin{equation}\label{codpeq1}
    \Vert \Tilde{\bm{\beta}}_{\mathrm{private},j} - \bm{\beta}_{0,j} \Vert_2 \lesssim \sqrt{\frac{q+\log n}{n}} + \sqrt{\frac{q\log(1/\delta) \log n}{n\epsilon^2}},
    \end{equation}
    with probability at least $1-CTn^{-1} - C_1e^{-q}$. 
\end{theorem}

Theorem \ref{codp} provides the estimation accuracy of the privacy-preserving version of \texttt{DSG-cqr}. The additional term, $O\left(\sqrt{\frac{q\log(1/\delta) \log n}{n\epsilon^2}}\right)$, arises from the noise added during each iteration. This term can be minimized by selecting a proper combination of $(\epsilon, \delta)$. In this case, increasing $(\epsilon, \delta)$ results in weaker privacy protection, hence indicating the existence of a potential  ``optimal privacy combination'': $\frac{\log(1/\delta)}{\epsilon^2} = O\left(\frac{q+\log n}{q\log n}\right)$, which represents the maximum allowable noise (for optimal privacy protection) while maintaining statistical accuracy.

\subsection{Feature-Distributed Inference for Quantile Regression}

To develop a more comprehensive statistical understanding of the estimator, a straightforward approach is to conduct statistical inferences, such as constructing confidence intervals for linear combinations of the regression coefficients. Under mild conditions, \cite{Koenker1978} demonstrate that the asymptotic covariance of the quantile regression coefficient relies on the population Hessian matrix $\mathbf{\bm{H}}_0 = \mathbb{E}(f_{\varepsilon|\bm{x}}(0)\bm{x}\bm{x}^\top)$ and $\bm{\Sigma}$. As suggested by \cite{He2023}, these two matrices can be estimated in the non-distributed setting by $\hat{\bm{\Sigma}} = \frac{1}{n} \sum_{i=1}^n \bm{x}_{i}\bm{x}_{i}^\top ~~ \text{and} ~~ \hat{\mathbf{\bm{H}}} = \frac{1}{nh} \sum_{i=1}^n K(\hat{\varepsilon}_i/h) \bm{x}_{i}\bm{x}_{i}^\top,$ which is known as the Powell's kernel-type estimator \citep{powell1988}, with $\hat{\varepsilon}_i$ denoting the estimated residuals. However, in the context of feature-distributed datasets, where each machine has limited access to global information, two challenges emerge: (1) the computation of estimated global residuals becomes infeasible, and (2) it is hard to calculate the off-diagonal components of $\hat{\bm{\Sigma}}$ and $\hat{\mathbf{\bm{H}}}$.

The first challenge is that the global residual cannot be directly calculated, as each machine can only access its parameters and raw data. In the \texttt{DSG-cqr} algorithm, it is noteworthy that each auxiliary variable $z_{ij}^{(t)}$ will, after sufficient iterations, approximate the product $\frac{1}{m}\sum_{j=1}^m \bm{x}_{ij}\bm{\beta}_j^{(t)}$. Thus, after $T$ steps of iteration ($T$ defined in Theorem \ref{th2}) we estimate the residual by defining the residual function 
\begin{equation}\label{est_res}
    \hat{\varepsilon}_i = \Tilde{\varepsilon}_i\left(z_{ij}^{(T)}\right) = y_i - mz_{ij}^{(T)}, ~~ i = 1,2,\cdots,n ~~ \text{and} ~~ j = 1,2,\cdots,m,
\end{equation}

The second challenge involves estimating the off-diagonal components, which requires access to raw data from multiple machines. While some studies have addressed covariance estimation in feature-distributed settings, they primarily focus on data publishing rules \citep{Govada2016} or assume shared memory across machines \citep{Hsieh2012}. To our knowledge, estimating cross-covariance for feature-distributed datasets within a connected network without raw data communication remains unresolved. Consequently, we impose the following block-diagonal assumption (abbreviated as BD).

BD) Matrices $\mathbf{\bm{H}}_0$ and $\bm{\Sigma}$ satisfy a block-independent structure across machines, that is, $\mathbf{\bm{H}}_0 = \text{diag}\{\mathbf{\bm{H}}_{0, 1}, \mathbf{\bm{H}}_{0, 2}, \cdots, \mathbf{\bm{H}}_{0, m}\}, \bm{\Sigma} = \text{diag}\{\bm{\Sigma}_1, \bm{\Sigma}_2, \cdots, \bm{\Sigma}_m\}$, where $\mathbf{\bm{H}}_{0, j}$ and $\bm{\Sigma}_j$ are both $p_j\times p_j$ matrices with $j = 1,\cdots, m$.

Condition BD) characterizes the independence structure across different machines. It allows for correlations among features within a single machine while assuming no correlations among features across different machines. Although more stringent, this condition is not contradictory to condition C5). A similar block-diagonal assumption is also considered in the decentralized, federated learning \citep{Valdeira2022}. Under BD), it is straightforward to have $(\mathbf{\bm{H}}_0^{-1})_j = \mathbf{\bm{H}}_{0, j}^{-1}$, where $(\mathbf{\bm{H}}_0^{-1})_j$ represents the $j$-th diagonal part of $\mathbf{\bm{H}}_0^{-1}$, which brings convenience to the inference. 

\begin{proposition}\label{prop1}
    Assume that the conditions C1) - C5) and BD) hold, and $h \asymp \left( \frac{q+\log n}{n} \right)^{\gamma}$ for any $\gamma \in [1/3, 1/2]$. Then, we have
    \begin{align*}
        & ~~~~\left\Vert \mathbf{\bm{H}}_{0, j}\left(\Tilde{\bm{\beta}}_j - \bm{\beta}_{0, j}\right) - \frac{1}{n} \sum_{i=1}^n  \left\{ \tau - \Bar{K}(-\varepsilon_{i}/h) \right\} \bm{x}_{ij}\right\Vert_2\\
        & \lesssim \frac{(q + \log n)^{1/2}(p + \log n)^{1/2}}{nh^{1/2}} + h\sqrt{\frac{p + \log n}{n}}.
    \end{align*}
\end{proposition}
Proposition \ref{prop1} can be viewed as a variant of the second part of Theorem 4.2 in \cite{He2023}, both of which depend on $n, p$, and $q$. It provides the theoretical guarantees for establishing the limiting distribution of the \texttt{DSG-cqr} estimator and its functionals. Theoretically, we suggest choosing the optimal bandwidth $h \asymp h_c\times\left( \frac{q + \log n}{n} \right)^{1/3},$ where $h_c$ is a constant that can be chosen from grid search and cross-validation in practice. This results in an optimal local Bahadur representation error of order $O\left(\frac{(q + \log n)^{1/3}(p + \log n)^{1/2}}{n^{5/6}}\right)$. Based on Proposition \ref{prop1}, we establish the statistical inference result for the \texttt{DSG-cqr} estimator.
    
\begin{theorem}\label{th3}
    Assume that the conditions in Proposition 1 hold. If $p_j = O(q),~ \forall j$, for any unit vector $\bm{v_j} \in \mathbb{R}^{p_j}$, we have
    \begin{equation}\label{th3eq1}
        \sup_{x \in \mathbb{R}} \left\vert \mathbb{P}\left(\sqrt{n}\langle\bm{v}_j,\Tilde{\bm{\beta}}_j - \bm{\beta}_{0,j}\rangle \leq x(\bm{v_j}^\top\bm{\sigma}_{h, j}^2\bm{v_j})^{1/2}  \right) - \Phi(x) \right\vert \lesssim \frac{(q + \log n)^{1/2}(p + \log n)^{1/2}}{(nh)^{1/2}} + n^{1/2}h^2,
    \end{equation}
    where $\bm{\sigma}_{h, j}^2 =  \mathbf{\bm{H}}_{0,j}^{-1}\mathbb{E}[\left\{ \Bar{K}(-\varepsilon/h)-\tau \right\}^2 \bm{x}_j\bm{x_j}^\top]\mathbf{\bm{H}}_{0,j}^{-1}$, $\Phi(\cdot)$ represents the standard normal distribution function. Specifically, by choosing $h \asymp \left( \frac{(q + \log n)^{1/5}(p + \log n)^{1/5}}{n^{2/5}} \right)$, $pq = o(n^{3/4})$, as $n \to \infty$, we have
    $$\frac{\sqrt{n} \langle\bm{v}_j,\Tilde{\bm{\beta}}_j - \bm{\beta}_{0,j}\rangle}{\left(\bm{v_j}^\top \mathbf{\bm{H}}_{0,j}^{-1} \Sigma_j \mathbf{\bm{H}}_{0,j}^{-1} \bm{v_j}\right)^{1/2}} \stackrel{d}{\to} N(0, \tau(1-\tau)).$$
\end{theorem}

In Theorem \ref{th3}, minimizing the normal approximation bounds \eqref{th3eq1} yields an optimal approximation error of order $O\left((q + \log n)^{2/5}(p + \log n)^{2/5}n^{-3/10}\right)$. Therefore, we requires $(pq)^{4/3}/n \to 0$ to ensure the normality of $\sqrt{n}\langle\bm{v}_j,\Tilde{\bm{\beta}}_j - \bm{\beta}_{0,j}\rangle$ for any unit vector $\bm{v}_j$ in the feature distributed setting. We present several scaling conditions necessary for normal approximation under random design for the quantile regression model, as outlined in Table \ref{Scal_Cond} for comparison, while local normality is considered for distributed datasets.

\begin{table}[h]\scriptsize
\setlength{\tabcolsep}{4pt}
\caption{Scaling conditions required for normal approximation under random design for the quantile loss function. The abbreviations used in the second column are defined as follows: ND: non-distributed; SD: sample-distributed; FD: feature-distributed.}
\label{Scal_Cond}
\begin{tabular}{cccc}
\toprule
\multicolumn{1}{c}{Literature} & \multicolumn{1}{c}{Design} & \multicolumn{1}{c}{Condition} & \multicolumn{1}{c}{Remark} \\ \midrule
\cite{Pan2020}                               & ND                          &  $p^3(\log n)^2 = o(n)$              &  Quantile regression                    \\
\cite{He2023}                               & ND                           & $p^{8/3} = o(n)$                      & \texttt{Conquer}             \\
\cite{Tan2021}                               & SD                            & $p = o(n^{2/5} \wedge N^{3/8})$                   & \texttt{Conquer}, $N$: Total sample size, $n$: local sample size                 \\
This paper                               & FD                           &   $(pq)^{4/3} = o(n)$                  &  \texttt{Conquer}, $p$: Total feature size, $q$: local feature size               \\ \bottomrule
\end{tabular}
\end{table}

Constructing the confidence interval for the regression coefficients requires estimating the asymptotic variance $\bm{\sigma}_{h, j}^2$. Under condition BD), we adopt the plug-in principle by replacing the components with its estimator, that is,
\begin{equation}\label{est_cov}
    \hat{\bm{\sigma}}_{h, j}^2 = \tau (1-\tau) \hat{\mathbf{\bm{H}}}_{h, j}^{-1} \hat{\bm{\Sigma}}_j \hat{\mathbf{\bm{H}}}_{h, j}^{-1},
\end{equation}
where $\hat{\mathbf{\bm{H}}}_{h, j} = \frac{1}{nh}\sum_{i=1}^n K(\hat{\varepsilon}_{ij}/h)\bm{x}_{ij}\bm{x}_{ij}^\top ~~ \mathrm{and} ~~ \hat{\bm{\Sigma}}_j = \frac{1}{n}\sum_{i=1}^n \bm{x}_{ij}\bm{x}_{ij}^\top,~~ j = 1, 2, \cdots, m$, and the estimated residual can be calculated within each machine through their local auxiliary variables, as in \eqref{est_res}.

\begin{remark}\label{est_covc}
    In a simpler case where the error term is independently and identically distributed (i.i.d.) and independent of the covariates, the covariance reduces to $\bm{\sigma}_{h, j}^2 = \frac{\tau(1-\tau)\bm{\Sigma}_j^{-1}}{f^2_{\varepsilon}(0)}$. Therefore, we only need to estimate the univariate density function at 0. The most commonly used kernel density estimator is $\hat{f}_{\varepsilon}(0) = \frac{1}{nh} \sum_{i=1}^n K(\hat{\varepsilon}_{ij}/h)$, which can be calculated within each machine according to \eqref{est_res}. This yields the estimated covariance $\hat{\bm{\sigma}}_{h, j}^2 = \tau(1-\tau)\hat{f}^{-2}_{\varepsilon}(0)\hat{\bm{\Sigma}}_j^{-1}$. However, it should be noted that the estimator of this form heavily relies on the assumption of unconditional error, making it less robust than \eqref{est_cov} in the presence of conditional heteroscedastic error.
\end{remark}

\begin{theorem}\label{th4}
    Assume that the conditions in Proposition \ref{prop1} hold, and let $h \asymp \left( \frac{q + \log n}{n} \right)^{1/3}$. If $p + \log n \lesssim n$, the estimator $\hat{\mathbf{\bm{H}}}_{h, j}$ satisfies
    $$\Vert \hat{\mathbf{\bm{H}}}_{h, j} - \mathbf{\bm{H}}_{0, j} \Vert_2 \lesssim \left(\frac{q + \log n}{n}\right)^{1/3} \bigvee \left(\frac{p + \log n}{n}\right)^{1/2}$$
    with probability tending to 1 as $n\to\infty$. Furthermore, we have $\frac{\hat{\bm{\sigma}}_{h, j}^2}{\bm{\sigma}_{h, j}^2} \stackrel{p}\to 1$.
\end{theorem}

Theorem \ref{th4} provides the estimation error of $\hat{\mathbf{\bm{H}}}_{h, j}$, and the consistency of the variance can be established as long as $p + \log n \lesssim n$. This ensures the validity of the inference procedure.

\section{Simulation Studies}\label{sec:sim}

In this section, we conduct simulations to evaluate the performance of the proposed \texttt{DSG-cqr} algorithm. As noted by \cite{He2023}, the \texttt{conquer} smoothing technique is robust to the choice of kernel functions, and therefore, we use the Gaussian kernel throughout this section. All simulations are performed using R 4.1.2 on a Linux platform with a 64-core AMD 7H12 (2.6GHz) CPU and 256GB of RAM. The R packages \texttt{quantreg} and \texttt{conquer} are loaded to obtain the standard quantile estimator and the \texttt{conquer} estimator, respectively.

\subsection{Feature-Distributed Quantile Regression}

We generate samples from model \eqref{qrmodel}, where the covariates $\bm{x}_i$ are i.i.d. sampled from the multivariate uniform distribution on the cube $\sqrt{3}\cdot [-1,1]^p$ with covariance matrix $\Sigma_x = [0.5^{|i-j|}]_{1\leq i, j \leq p}$. The true regression coefficients $\bm{\beta}_0 = \bm{b}_1 \bm{b}_2$, where each element of $\bm{b}_1$ are i.i.d. drawn from $\{-1, 1\}$ with probabilities $\mathbb{P}(b = -1) = \mathbb{P}(b = 1) = 1/2$, and each element of $\bm{b}_2$ is drawn from a uniform distribution $U(1, 2)$. For the error term, we consider the following two scenarios: Scenario 1: (Homoscedastic Case) $\varepsilon_{i}=e_i-q_e(\tau)$; and Scenario 2: (Conditional Heteroscedastic Case) $\varepsilon_{i}= (1+0.25x_{i1})(e_i - q_e(\tau))$, with $q_e(\tau)$ being the $\tau$-th quantile of the distribution function of $e_i$. We also consider two types of $e_i$: the standard normal distribution $N(0, 1)$, and the $t$-distribution with 5 degrees of freedom $t(5)$. We normalize the $t$-distribution with constant $\sqrt{3/5}$ so that both types of $e_i$ share the same variance. We consider $p = 60$, $m \in \{6, 15\}$, $\tau = 0.25$, and let each $p_j$ equal to $p/m$. For the mixing matrix $\bm{W} = [w_{ij}]$, we consider the Metropolis-Hastings weight matrix defined as
\begin{eqnarray*}
    w_{ij}    \left\{
    \begin{array}{ccl}
            \left(\max\{\mathrm{deg}(i), \mathrm{deg}(j)\} +1\right)^{-1}, && \mathrm{if}~~(i, j)\in \mathcal{E},   \\
            0, && \mathrm{if}~~(i, j)\notin \mathcal{E}~~\mathrm{and}~~i \neq j, \\
            1 - \sum_{k\in \mathcal{V}} w_{ik}, && \mathrm{if}~~i = j,
    \end{array}\right.
\end{eqnarray*}
where $\mathrm{deg}(i)$ refers to the degree of machine $i$. We set the size of the edges to $0.5m(m-1)\pi_W$ with $\pi_W = 0.5$, and each edge is randomly selected from all possible edges. Throughout this subsection, we choose $h = 1.5 \left(\frac{(p + \log n)(1.5 \phi(\Phi^{-1}(\tau))^2)}{n(2\Phi^{-1}(\tau)^2 + 1)} \right)^{1/3}$, where $\phi(\cdot)$ and $\Phi(\cdot)$ represent the p.d.f. and c.d.f. of the standard normal distribution, respectively. The multiplier term related to $\tau$ is selected according to the rule-of-thumb principle \citep{Tan2021}.

\begin{table}[h]\scriptsize
\setlength{\tabcolsep}{6pt}
\caption{Mean (standard deviation) testing error over 100 simulation runs.}
\label{CVERR}
\begin{tabular}{cccccccc}
\toprule

\multirow{2}{*}{$\tau$} & \multirow{2}{*}{Method} & \multicolumn{3}{c}{$N(0, 1)$}                & \multicolumn{3}{c}{$t(5)$}                   \\ \cline{3-8} 
                     &                         & $n=5000$        & $n=10000$        & $n=20000$       & $n=5000$        & $n=10000$        & $n=20000$       \\ \midrule
\multicolumn{8}{l}{\textbf{Panel (a)}: Homoscedastic case}                                                                                                      \\ 
\multirow{6}{*}{0.25} & \texttt{DSG-cqr}                & 3.196(0.107)  & 1.601(0.043)  & 0.798(0.014)  & 3.017(0.129)  & 1.515(0.048)  & 0.755(0.021)  \\
                      & \texttt{DSG-cqr (PP)}                & 3.218(0.114)  & 1.597(0.045)  & 0.805(0.015)  & 3.044(0.154)  & 1.518(0.048)  & 0.765(0.023)  \\
                      & \texttt{glb-qr}                   & 3.201(0.107)  & 1.602(0.043)  & 0.798(0.014)  & 3.022(0.129)  & 1.516(0.048)  & 0.755(0.021)  \\
                      & \texttt{glb-cqr}                & 3.228(0.119)  & 1.598(0.035)  & 0.798(0.015)  & 3.033(0.134)  & 1.507(0.050)  & 0.755(0.018)  \\
                      & \texttt{iso-qr}                   & 43.725(4.649) & 21.833(1.835) & 11.096(0.931) & 44.650(4.233)  & 22.259(1.772) & 10.973(0.988) \\
                      & \texttt{iso-cqr}                & 36.962(2.916) & 18.030(1.512) & 9.070(0.698)  & 36.014(2.863) & 18.132(1.606) & 9.252(0.682)  \\
\multirow{6}{*}{0.5}  & \texttt{DSG-cqr}                & 4.021(0.134)  & 2.003(0.052)  & 1.000(0.019)  & 3.682(0.155)  & 1.837(0.055)  & 0.924(0.018)  \\
                      & \texttt{DSG-cqr (PP)}                & 4.118(0.144)  & 2.047(0.042)  & 1.055(0.022)  & 3.699(0.164)  & 1.858(0.058)  & 0.935(0.020)  \\
                      & \texttt{glb-qr}                   & 4.027(0.135)  & 2.005(0.052)  & 1.000(0.019)  & 3.688(0.154)  & 1.838(0.056)  & 0.925(0.018)  \\
                      & \texttt{glb-cqr}                & 4.041(0.127)  & 2.002(0.048)  & 1.001(0.019)  & 3.706(0.146)  & 1.849(0.054)  & 0.923(0.020)  \\
                      & \texttt{iso-qr}                   & 45.661(3.489) & 23.137(1.748) & 11.456(0.928) & 45.491(4.108) & 22.843(1.877) & 11.541(0.923) \\
                      & \texttt{iso-cqr}                & 46.060(4.084) & 22.524(1.988) & 11.426(0.979) & 46.088(3.973) & 22.940(1.764) & 11.340(0.840) \\
\multirow{6}{*}{0.75} & \texttt{DSG-cqr}                & 3.230(0.108)  & 1.600(0.038)  & 0.797(0.016)  & 3.036(0.150)  & 1.513(0.046)  & 0.756(0.019)  \\
                      & \texttt{DSG-cqr (PP)}                & 3.238(0.114)  & 1.597(0.042)  & 0.802(0.017)  & 3.044(0.154)  & 1.518(0.048)  & 0.755(0.021)  \\
                      & \texttt{glb-qr}                   & 3.235(0.108)  & 1.602(0.038)  & 0.798(0.016)  & 3.042(0.148)  & 1.514(0.046)  & 0.756(0.019)  \\
                      & \texttt{glb-cqr}                & 3.222(0.123)  & 1.594(0.038)  & 0.796(0.014)  & 3.066(0.153)  & 1.510(0.048)  & 0.756(0.020)  \\
                      & \texttt{iso-qr}                   & 43.925(3.830) & 22.052(1.990) & 11.041(0.901) & 45.072(3.915) & 22.501(1.800) & 11.200(0.992) \\
                      & \texttt{iso-cqr}                & 36.900(3.096) & 18.214(1.574) & 9.172(0.800)  & 36.361(3.103) & 18.136(1.255) & 9.055(0.713)  \\  \midrule
\multicolumn{8}{l}{\textbf{Panel (b)}: Heteroscedastic case}                                                                                                      \\
\multirow{6}{*}{0.25} & \texttt{DSG-cqr}                & 3.218(0.124)  & 1.597(0.042)  & 0.795(0.014)  & 3.044(0.154)  & 1.518(0.048)  & 0.755(0.020)  \\
                      & \texttt{DSG-cqr (PP)}                & 3.238(0.130)  & 1.605(0.045)  & 0.803(0.014)  & 3.052(0.160)  & 1.520(0.050)  & 0.754(0.019)  \\
                      & \texttt{glb-qr}                   & 3.223(0.124)  & 1.599(0.042)  & 0.796(0.014)  & 3.048(0.155)  & 1.518(0.048)  & 0.755(0.020)  \\
                      & \texttt{glb-cqr}                & 3.206(0.121)  & 1.593(0.044)  & 0.798(0.016)  & 3.030(0.147)  & 1.508(0.049)  & 0.754(0.019)  \\
                      & \texttt{iso-qr}                   & 43.970(3.914) & 21.791(1.976) & 10.909(0.974) & 43.455(4.045) & 22.188(1.947) & 11.103(0.973) \\
                      & \texttt{iso-cqr}                & 36.348(3.427) & 18.103(1.351) & 9.124(0.767)  & 36.201(3.000) & 18.187(1.528) & 9.111(0.764)  \\
\multirow{6}{*}{0.5}  & \texttt{DSG-cqr}                & 4.001(0.131)  & 2.005(0.055)  & 1.001(0.018)  & 3.719(0.176)  & 1.847(0.055)  & 0.922(0.021)  \\
                      & \texttt{DSG-cqr (PP)}                & 4.022(0.134)  & 2.007(0.056)  & 1.010(0.019)  & 3.740(0.184)  & 1.877(0.058)  & 0.933(0.020)  \\
                      & \texttt{glb-qr}                   & 4.007(0.132)  & 2.007(0.056)  & 1.001(0.018)  & 3.723(0.176)  & 1.849(0.055)  & 0.922(0.021)  \\
                      & \texttt{glb-cqr}                & 4.005(0.147)  & 2.010(0.052)  & 1.000(0.018)  & 3.714(0.166)  & 1.849(0.055)  & 0.921(0.022)  \\
                      & \texttt{iso-qr}                   & 45.884(3.870) & 22.593(1.847) & 11.417(0.786) & 45.468(3.660) & 22.804(1.645) & 11.505(0.920) \\
                      & \texttt{iso-cqr}                & 45.701(3.380) & 22.753(1.796) & 11.473(0.916) & 46.155(3.761) & 22.800(1.577) & 11.549(0.852) \\
\multirow{6}{*}{0.75} & \texttt{DSG-cqr}                & 3.200(0.103)  & 1.597(0.045)  & 0.796(0.015)  & 3.066(0.149)  & 1.506(0.056)  & 0.754(0.020)  \\
                      & \texttt{DSG-cqr (PP)}                & 3.218(0.125)  & 1.598(0.047)  & 0.795(0.014)  & 3.074(0.154)  & 1.518(0.060)  & 0.758(0.022)  \\
                      & \texttt{glb-qr}                   & 3.206(0.104)  & 1.598(0.045)  & 0.796(0.015)  & 3.071(0.149)  & 1.508(0.056)  & 0.754(0.020)  \\
                      & \texttt{glb-cqr}                & 3.196(0.110)  & 1.599(0.043)  & 0.799(0.015)  & 3.039(0.164)  & 1.513(0.053)  & 0.753(0.018)  \\
                      & \texttt{iso-qr}                   & 44.794(3.915) & 21.972(1.931) & 10.987(0.774) & 44.697(3.277) & 22.027(1.778) & 11.096(0.927) \\
                      & \texttt{iso-cqr}                & 36.249(3.287) & 17.841(1.395) & 9.202(0.722)  & 36.557(2.727) & 18.151(1.577) & 9.083(0.717)  \\ \bottomrule
\end{tabular}
\end{table}

\begin{figure}[h]
  \centering
  \includegraphics[width=\textwidth]{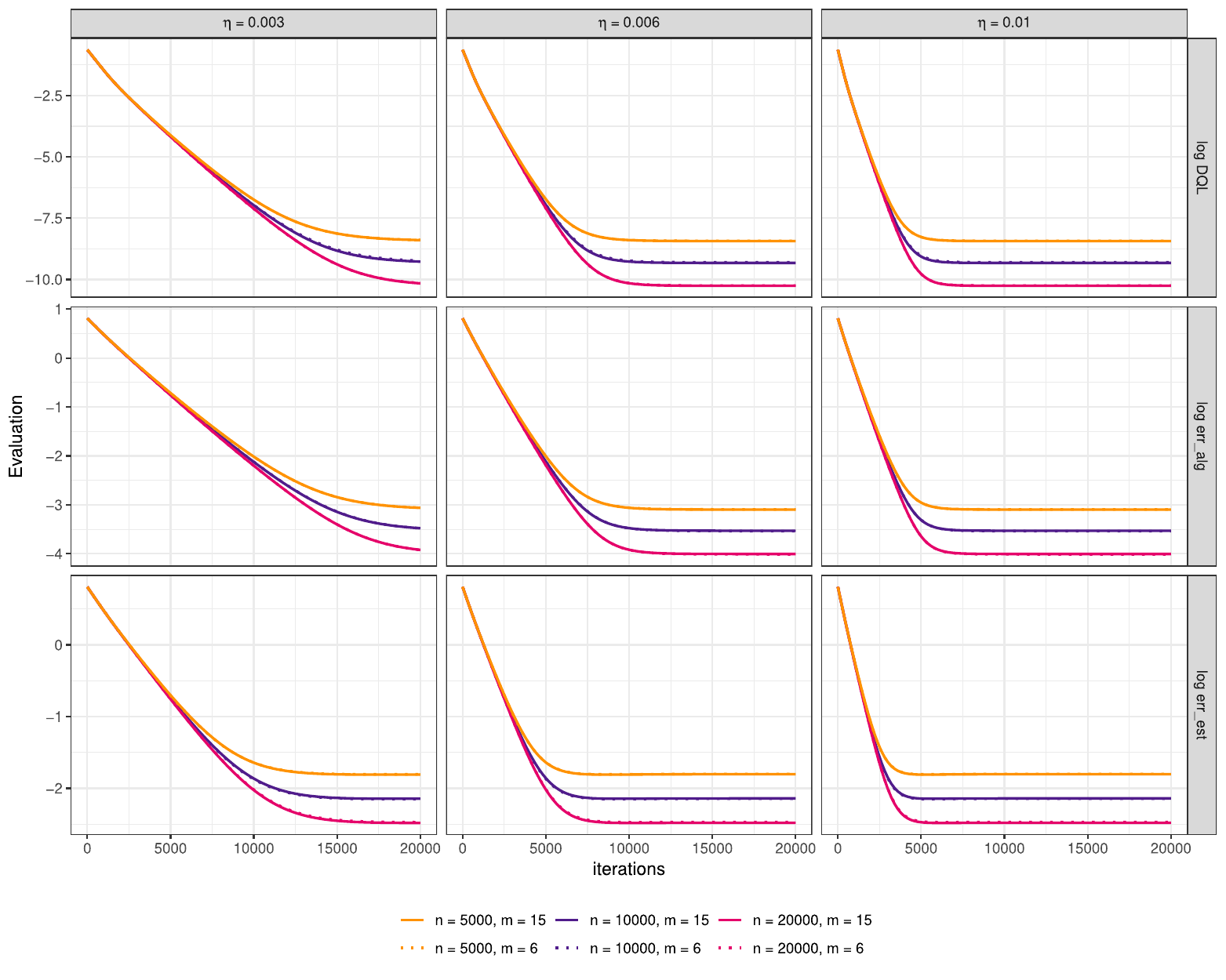}
  \caption{Convergence path with $n \in \{ 5000, 10000, 20000\}, m \in \{ 6, 15\}, \eta \in \{ 0.003, 0.006, 0.01\}, \tau = 0.25$.}
  \label{SimNM2}
\end{figure}

We conduct our simulation by randomly splitting the data into the training and testing sets, with 90\% allocated to the training data and 10\% to the testing data. Several estimation procedures are then carried out based on the training data. We report the testing error averaged over 100 simulation runs, that is, $\frac{1}{100n_{\text{test}}} \sum_{b=1}^{100} \sum_{i\in{\mathcal{S}_\text{test}}} \rho_\tau \left(y_i - \sum_{j=1}^m \bm{x}_{ij}^\top\Tilde{\bm{\beta}}_{\text{train}}^{[b]}\right)$, where $\Tilde{\bm{\beta}}_{\text{train}}^{[b]}$ is the estimator based on the training set at the $b$-th replication, $\mathcal{S}_\text{test}$ is the index set of the testing data, and $n_{\text{test}}$ represents its cardinality. We compare \texttt{DSG-cqr} and its privacy-preserving counterpart \texttt{DSG-cqr (PP)} with four competing methods: (i) quantile estimator based on the global datasets, \texttt{glb-qr}; (ii) \texttt{conquer} estimator based on the global datasets, \texttt{glb-cqr}; (iii) quantile estimator based on the datasets within the 1-st machine, \texttt{iso-qr}; and (iv) \texttt{conquer} estimator based on the datasets within the 1-st machine, \texttt{iso-qr}. Specifically, we set the overall privacy level $\Bar{\epsilon} = 0.5$ for \texttt{DSG-cqr (PP)},   within each iteration as $\epsilon^{-1}\sqrt{2\log(1.25/\delta)} = \sqrt{\frac{q+\log n}{q\log n}}$, and the $l_2$ sensitivity is selected according to Remark \ref{rmk_pv}. Table \ref{CVERR} reports the results of testing error (standard deviation) with $n \in \{5000, 10000, 20000\}, m = 15$. It shows that when $\tau=0.25$, the testing error of \texttt{DSG-cqr} closely aligns with that of \texttt{glb-cqr} but significantly outperforms \texttt{iso-cqr}, irrespective of whether the error term exhibits heteroscedasticity or heavy-tailed characteristics. This can be explained by the insufficient feature information and potential bias caused by the model misspecification. Similar patterns can also be observed for $\tau \in \{0.5, 0.75\}$. Due to its smoothed nature, \texttt{conquer} performs slightly better than the ordinary quantile estimator, particularly in cases with small sample sizes or estimating non-median quantile levels. Additionally, the privacy-preserving algorithm \texttt{DSG-cqr (PP)} provides almost the same results as \texttt{DSG-cqr}.

\noindent\textbf{Convergence Path.} ~ We report the convergence paths of three quantities during the updating steps for the proposed \texttt{DSG-cqr} algorithm: (i) the difference of the quantile loss (DQL), defined as $\text{DQL}^{(t)} = |\hat{Q}_\tau(\bm{\beta}^{(t)}) - \hat{Q}_\tau(\bm{\beta}^*)|$, where $\bm{\beta}^*$ is the quantile estimator based on the global datasets; (ii) the estimation error, measured by $\Vert\bm{\beta}^{(t)} - \bm{\beta}_0 \Vert_2$; and (iii) the algorithm error, measured by $\Vert\bm{\beta}^{(t)} - \bm{\beta}^* \Vert_2$. We consider different step sizes $\eta \in \{0.003, 0.006, 0.01\}$ and $m \in \{6, 15\}$, with the mixing round ${\kappa_0}$ set to 1. Figure \ref{SimNM2} presents the convergence paths for these three quantities after logarithmic transformations averaged over 100 simulation runs with $\tau = 0.25$, and results for $\tau \in \{0.5, 0.75\}$ are available in Appendix. From Figure \ref{SimNM2}, it can be observed that \texttt{DSG-cqr} nearly achieves a linear convergence rate and converges more rapidly with a larger step size. The estimation error decreases more quickly and yields more accurate results with larger sample sizes. Notably, the estimation error converges in fewer iterations compared to the algorithm error. This phenomenon can be explained by the dominance of statistical error, i.e., $\Vert\bm{\beta}^{(t)} - \bm{\beta}_0 \Vert_2 \leq \Vert\bm{\beta}^{(t)} - \bm{\beta}^* \Vert_2 + \Vert\bm{\beta}^* - \bm{\beta}_0 \Vert_2$, where the convergence of $\Vert\bm{\beta}^{(t)} - \bm{\beta}_0 \Vert_2$ only requires $\Vert\bm{\beta}^{(t)} - \bm{\beta}^* \Vert_2 \lesssim \Vert\bm{\beta}^* - \bm{\beta}_0 \Vert_2$, which does not guarantee the convergence of $\Vert\bm{\beta}^{(t)} - \bm{\beta}^* \Vert_2$. The parameter $\pi_W$ of the Metropolis-Hastings network is fixed, so $m$ has a minor effect on the overall convergence, as Figure \ref{SimNM2} shows. In the remainder of this subsection, we will conduct simulations to demonstrate the impact of different algorithm-related parameters on the finite sample performance of \texttt{DSG-cqr}. To save space, we only report results with $\tau = 0.25$ and Scenario 2 with $N(0,1)$ errors.

\begin{figure}[h]
  \centering
  \includegraphics[width=\textwidth]{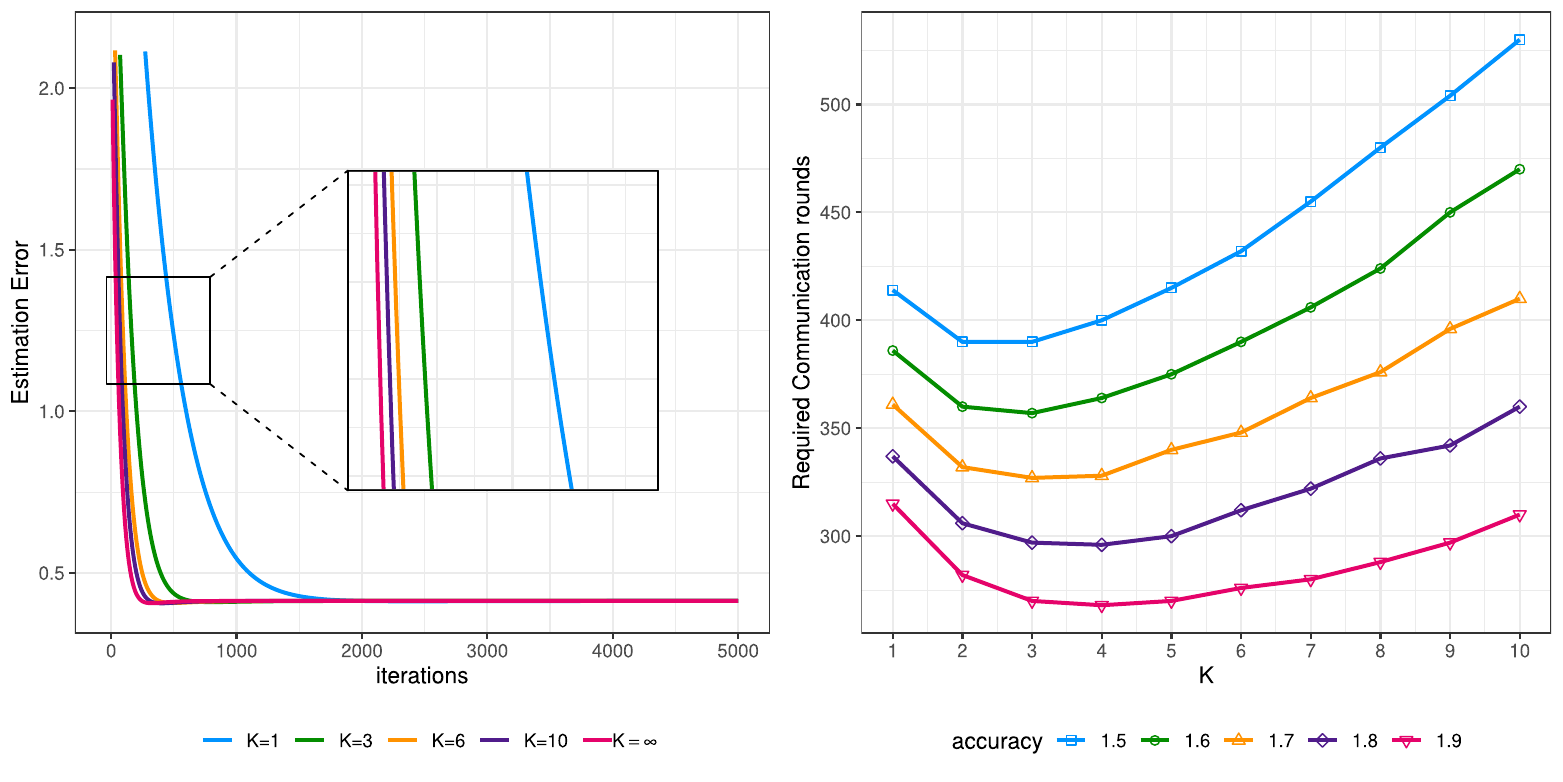}
  \caption{The left panel shows the estimation error with different mixing rounds ${\kappa_0} \in\{1, 3, 6, 10\}$, and an additional ${\kappa_0} = \infty$ that refers to the centralized case. The right panel presents the required communication rounds with different accuracy values of ${\kappa_0}$. In both panels, $n = 800$, $p = 60$, $m = 15$, $\eta = 0.1$, and the network is a fixed Line structure.}
  \label{SimK}
\end{figure}

\noindent\textbf{Impact of Mixing Rounds.} ~ We perform a simulation with varying numbers of mixing rounds ${\kappa_0}$ to illustrate the potential trade-off between required communication rounds and accuracy for \texttt{DSG-cqr}. We use a slightly larger step size $\eta = 0.1$ and consider the Line network structure such that the optimal ${\kappa_0}$ will be more likely to be greater than 1 rather than equal to 1. We set $n = 800, m = 15$, and other settings are the same as the previous part of this subsection. The results are demonstrated in Figure \ref{SimK}, where ${\kappa_0} = \infty$ represents the centralized case used as a benchmark. It can be seen from the left panel of Figure \ref{SimK} that more mixing rounds lead to a faster convergence through consensus improvement. However, such an improvement gradually diminishes as ${\kappa_0}$ increases. The right panel shows the required communication rounds under a fixed accuracy of the loss function with different ${\kappa_0}$. It demonstrates that higher accuracy necessitates more communication rounds. Additionally, different values of ${\kappa_0}$ result in varying communication rounds needed to reach the same level of accuracy. For example, $\kappa_{0,\mathrm{opt}} = 2$ is required to achieve an accuracy of 1.5, while $\kappa_{0,\mathrm{opt}} = 3$ is necessary for accuracies of 1.7 and 1.8. These findings corroborate the impact of ${\kappa_0}$ as outlined in Theorem \ref{th1} and confirm the existence of $\kappa_{0,\mathrm{opt}}$ as established in Theorem \ref{th2}.

\begin{figure}[h]
  \centering
  \includegraphics[width=0.66\textwidth]{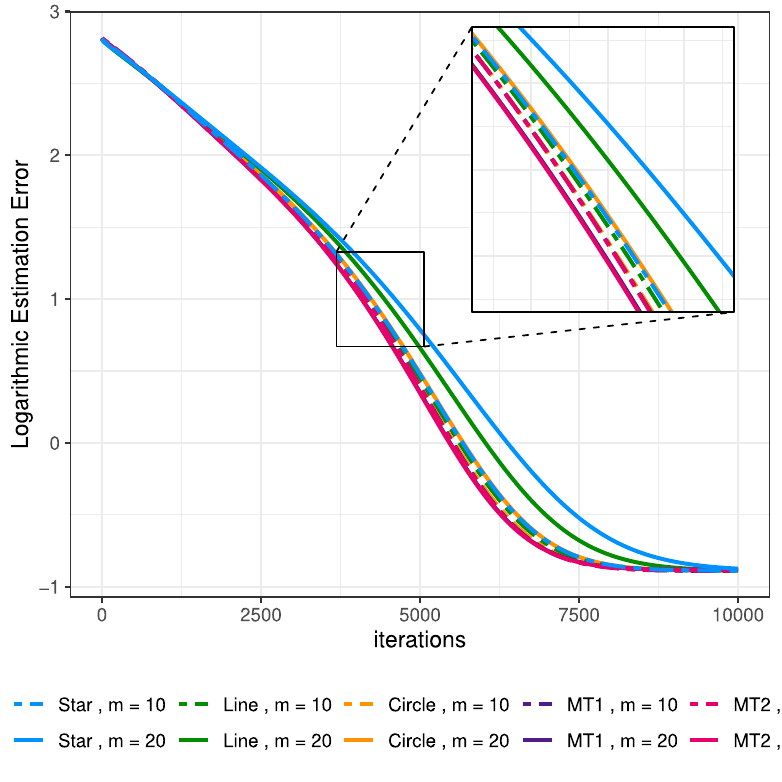}
  \caption{The convergence paths of different network structures.}
  \label{SimW}
\end{figure}

\noindent\textbf{Impact of Network Topology.} ~ We conduct a simulation to compare the effects of network topology for \texttt{DSG-cqr}. In this simulation, the mixing rounds are fixed at ${\kappa_0} = 2$, and we compare the following network structures: Star, Line, Circle, Metropolis-Hastings with $\pi_{W} = 0.4$ (MT1) and $\pi_{W} = 0.6$ (MT2). The convergence paths are shown in Figure \ref{SimW}, where the logarithmic algorithm error is reported for better illustration. As seen therein, with a fixed number of machines $m$, the convergence speeds are ranked as  Circle $>$ Line $>$ Star. This ranking can be explained by the relative sparsity of these networks, specifically, $\alpha_{\text{Circle}} < \alpha_{\text{Line}} < \alpha_{\text{Star}}$, as demonstrated by simple calculations. Additionally, the number of machines indirectly affects the convergence of \texttt{DSG-cqr} through network sparsity: a larger $m$ leads to a sparser network. These simulation results validate the impact of network structure on the convergence properties of \texttt{DSG-cqr}, as established in Theorem \ref{th1}.

\subsection{Feature-Distributed Confidence Interval Construction}

In this section, we conduct several simulations to illustrate the performance of the proposed confidence interval construction method. To distinguish between different covariance estimators, we use \texttt{(hr)} and \texttt{(hs)}, abbreviated as heteroscedasticity robust and heteroscedasticity sensitive, referring to the estimators in \eqref{est_cov} and Remark \ref{est_covc}, respectively. For comparing purposes, we also present simulation results for the normal approximation confidence interval (CI) generated by \texttt{conquer} \citep{Fernandes2021}, as well as for the ordinary quantile estimator \citep{Koenker1978}, in both global and isolated cases. We set $p = 30$, $n = 20000$, and $m \in \{6, 15\}$. The bandwidth is chosen as $h = 0.5 \left(\frac{(p + \log n)(1.5 \phi(\Phi^{-1}(\tau))^2)}{n(2\Phi^{-1}(\tau)^2 + 1)} \right)^{1/3}$. The design matrix is generated with a block-diagonal structure, $\Sigma_x^{'} = [0.5^{|i-j|}]$, if $i, j$ come from the same machine. Other details of the data generation process are the same as in Section 3.1. Two performance metrics are considered: the average empirical coverage probability (AECP) and the average width (AW) of the confidence interval for the first coefficient within the first and second machines. Both metrics are computed as the mean over 1000 simulation runs.

\begin{table}[h]\scriptsize
\setlength{\tabcolsep}{6pt}
\caption{Width of the confidence interval (empirical coverage probabilities) averaged over 1000 simulation runs within the 1st machine.}
\label{Sim_CI1}
\begin{tabular}{ccccccc}
\toprule
\multirow{2}{*}{Methods}  & \multicolumn{3}{c}{$N(0,1)$}                 & \multicolumn{3}{c}{$t(5)$}                     \\& $\tau = 0.25$   & $\tau = 0.5$    & $\tau = 0.75 $  & $\tau = 0.25$   & $\tau = 0.5$    & $\tau = 0.75$   \\ \midrule
                          \multicolumn{7}{l}{\textbf{Panel (a)}: Homoscedastic case} \\
\texttt{DSG-cqr(hr, m=6)}            & 4.259(0.950)  & 3.980(0.967) & 4.260(0.946) & 3.744(0.948) & 3.292(0.961) & 3.739(0.959) \\
 \texttt{DSG-cqr(hs, m=6)}            & 4.240(0.947) & 3.974(0.966)  & 4.242(0.948) & 3.726(0.949) & 3.284(0.962) & 3.723(0.960) \\
 \texttt{DSG-cqr(hr, m=15)}            & 4.230(0.955)  & 3.960(0.963) & 4.226(0.945) & 3.724(0.955) & 3.263(0.966) & 3.728(0.961) \\
 \texttt{DSG-cqr(hs, m=15)}            & 4.222(0.953) & 3.955(0.963)  & 8.220(0.944) & 3.714(0.953) & 3.262(0.967) & 3.715(0.962) \\
 \texttt{glb-qr}                & 4.205(0.936) & 3.959(0.953) & 4.207(0.938) & 3.685(0.943) & 3.275(0.955) & 3.697(0.947) \\
 \texttt{glb-cqr} & 4.283(0.950)   & 4.031(0.956)  & 4.281(0.948)  & 3.748(0.954)  & 3.318(0.955)  & 3.747(0.953)  \\
 \texttt{iso-cqr (m=6)} & 46.561(0.941) & 43.198(0.944) & 46.320(0.933)  & 46.697(0.944) & 43.343(0.959) & 46.450(0.940)   \\
 \texttt{iso-qr (m=6)} & 47.108(0.943) & 43.355(0.945) & 46.822(0.936) & 47.176(0.945) & 43.497(0.957) & 46.951(0.939) \\
 \texttt{iso-cqr (m=15)} & 48.987(0.943) & 45.683(0.955) & 49.241(0.945) & 48.870(0.951)  & 45.420(0.935)  & 49.071(0.941) \\
 \texttt{iso-qr (m=15)} & 49.329(0.949) & 45.811(0.960)  & 49.637(0.951) & 49.252(0.958) & 45.553(0.935) & 49.447(0.947) \\ \midrule
 \multicolumn{7}{l}{\textbf{Panel (b)}: Heteroscedastic case} \\
 \texttt{DSG-cqr(hr, m=6)}            & 3.865(0.955)  & 3.624(0.964) & 3.865(0.945) & 3.393(0.946) & 3.013(0.960) & 3.394(0.941) \\
 \texttt{DSG-cqr(hs, m=6)}            & 3.978(0.963) & 3.734(0.971)  & 3.978(0.958) & 3.495(0.949) & 3.098(0.962) & 3.496(0.945) \\
 \texttt{DSG-cqr(hr, m=15)}            & 3.836(0.949)  & 3.601(0.951) & 3.835(0.947) & 3.372(0.937) & 2.985(0.958) & 3.375(0.940) \\
 \texttt{DSG-cqr(hs, m=15)}            & 3.964(0.952) & 3.718(0.958)  & 3.958(0.950) & 3.479(0.946) & 3.076(0.971) & 3.480(0.951) \\
 \texttt{glb-qr}                & 3.945(0.953) & 3.721(0.959) & 3.945(0.948) & 3.466(0.951) & 3.091(0.962) & 3.466(0.946) \\
 \texttt{glb-cqr} & 3.969(0.943)  & 3.666(0.940)   & 3.973(0.937)  & 3.490(0.932)   & 3.044(0.944)  & 3.496(0.952)  \\
 \texttt{iso-cqr (m=6)} & 46.903(0.735) & 43.248(0.954) & 46.583(0.737) & 46.550(0.804)  & 43.394(0.954) & 46.375(0.823) \\
 \texttt{iso-qr (m=6)} & 47.400(0.740)    & 43.393(0.960)  & 47.077(0.744) & 47.035(0.811) & 43.539(0.955) & 46.898(0.831) \\
 \texttt{iso-cqr (m=15)} & 49.165(0.744) & 45.708(0.944) & 48.921(0.759) & 49.319(0.836) & 45.342(0.952) & 49.042(0.811) \\
 \texttt{iso-qr (m=15)} & 49.564(0.748) & 45.820(0.945)  & 49.251(0.764) & 49.661(0.838) & 45.478(0.954) & 49.381(0.820) \\ \hline
\end{tabular}
\end{table}

\begin{table}[h]\scriptsize
\setlength{\tabcolsep}{6pt}
\caption{Width of the confidence interval (empirical coverage probabilities) averaged over 1000 simulation runs within the 2nd machine.}
\label{Sim_CI2}
\begin{tabular}{ccccccc}
\toprule
\multirow{2}{*}{Methods} & \multicolumn{3}{c}{$N(0,1)$}                 & \multicolumn{3}{c}{$t(5)$}                     \\

& $\tau = 0.25$   & $\tau = 0.5$    & $\tau = 0.75 $  & $\tau = 0.25$   & $\tau = 0.5$    & $\tau = 0.75$   \\ \midrule
\multicolumn{7}{l}{\textbf{Panel (a)}: Homoscedastic case} \\
 \texttt{DSG-cqr(hr, m=6)}            & 4.265(0.947)  & 3.984(0.963) & 4.259(0.954) & 3.741(0.953) & 3.292(0.956) & 3.738(0.948) \\
 \texttt{DSG-cqr(hs, m=6)}            & 4.240(0.945) & 3.974(0.961)  & 4.241(0.956) & 3.727(0.952) & 3.285(0.954) & 3.722(0.946) \\
 \texttt{DSG-cqr(hr, m=15)}            & 4.234(0.951)  & 3.957(0.957) & 4.228(0.958) & 3.727(0.957) & 3.265(0.963) & 3.717(0.955) \\
 \texttt{DSG-cqr(hs, m=15)}            & 4.222(0.951) & 3.955(0.958)  & 4.220(0.953) & 3.713(0.961) & 3.263(0.964) & 3.714(0.958) \\
 \texttt{glb-qr}                & 4.205(0.936) & 3.959(0.953) & 4.207(0.938) & 3.685(0.943) & 3.275(0.955) & 3.697(0.947) \\
 \texttt{glb-cqr} & 4.283(0.950)   & 4.031(0.956)  & 4.281(0.948)  & 3.748(0.954)  & 3.318(0.955)  & 3.747(0.953)  \\
 \texttt{iso-cqr (m=6)} & 46.734(0.944) & 43.206(0.939) & 46.113(0.942) & 46.671(0.946) & 43.274(0.946) & 46.719(0.941) \\
 \texttt{iso-qr (m=6)} & 47.229(0.950)  & 43.357(0.939) & 46.597(0.945) & 47.172(0.950)  & 43.430(0.946)  & 47.231(0.944) \\
 \texttt{iso-cqr (m=15)} & 49.112(0.944) & 45.715(0.942) & 49.103(0.946) & 48.740(0.950)   & 45.534(0.953) & 48.818(0.939) \\
 \texttt{iso-qr (m=15)} & 49.510(0.945)  & 45.837(0.946) & 49.462(0.952) & 49.102(0.952) & 45.659(0.952) & 49.220(0.934) \\ \midrule
 \multicolumn{7}{l}{\textbf{Panel (b)}: Heteroscedastic case} \\
 \texttt{DSG-cqr(hr, m=6)}            & 3.995(0.956)  & 3.742(0.958) & 3.994(0.950) & 3.509(0.952) & 3.104(0.963) & 3.515(0.960) \\
 \texttt{DSG-cqr(hs, m=6)}            & 3.978(0.951) & 3.734(0.956)  & 3.977(0.952) & 3.494(0.952) & 3.097(0.963) & 3.495(0.953) \\
 \texttt{DSG-cqr(hr, m=15)}            & 3.969(0.958)  & 3.719(0.965) & 3.963(0.964) & 3.482(0.954) & 3.077(0.961) & 3.485(0.950) \\
 \texttt{DSG-cqr(hs, m=15)}            & 3.963(0.958) & 3.716(0.966)  & 3.959(0.961) & 3.479(0.953) & 3.075(0.961) & 3.480(0.952) \\
 \texttt{glb-qr}                & 3.945(0.953) & 3.721(0.959) & 3.945(0.948) & 3.466(0.951) & 3.091(0.962) & 3.466(0.946) \\
 \texttt{glb-cqr} & 3.969(0.943)  & 3.666(0.940)   & 3.973(0.937)  & 3.490(0.932)   & 3.044(0.944)  & 3.496(0.952)  \\
 \texttt{iso-cqr (m=6)} & 46.603(0.944) & 43.191(0.938) & 46.518(0.948) & 46.725(0.941) & 43.129(0.948) & 46.616(0.945) \\
 \texttt{iso-qr (m=6)} & 47.096(0.945) & 43.314(0.940)  & 46.988(0.952) & 47.218(0.942) & 43.287(0.953) & 47.140(0.952)  \\
 \texttt{iso-cqr (m=15)} & 49.349(0.938) & 45.808(0.932) & 48.814(0.943) & 49.161(0.936) & 45.665(0.958) & 49.133(0.942) \\
 \texttt{iso-qr (m=15)} & 49.733(0.941) & 45.925(0.932) & 49.211(0.949) & 49.544(0.939) & 45.783(0.962) & 49.458(0.945) \\ \bottomrule
\end{tabular}
\end{table}

The results for the first and second machines are presented in Tables \ref{Sim_CI1} and \ref{Sim_CI2}, respectively. It can be observed that the isolated methods produce markedly wider confidence intervals compared to the global methods, likely due to the failure of error density estimation with insufficient features. In terms of coverage probability, the isolated methods exhibit notable coverage distortion, particularly at non-median quantiles under conditional heteroscedastic errors.

\section{Empirical Illustration}\label{sec:emp}

This section illustrates the usefulness of our method by considering the Communities and Crime dataset from the UCI Machine Learning Repository (\url{https://archive.ics.uci.edu/dataset/183/communities+and+crime}). This dataset integrates community-level data from the 1990 United States Census (USC), the 1990 United States Law Enforcement Management and Administrative Statistics (LEMAS) survey, and crime data from the 1995 Federal Bureau of Investigation Uniform Crime Reporting (FBI-UCR). The objective is to investigate the relationship between community characteristics and the total number of violent crimes per $10^5$ population across different quantile levels. Given that the features are collected from various sources (e.g., social, economic, infrastructure), it is natural to assume that they are stored in different departments with restrictions on raw data sharing. Accordingly, we group the features based on the criteria outlined in Table \ref{Emp_Group_Tab}.

\begin{table}[h]\scriptsize
\setlength{\tabcolsep}{2pt}
\caption{Grouping rules for features of the Communities and Crime dataset.}
\label{Emp_Group_Tab}
\begin{tabular}{ccc}
\toprule
\multicolumn{1}{c}{No. of Vars.} & \multicolumn{1}{c}{Description} & \multicolumn{1}{c}{Feature} \\ \midrule
6                       & Public facilities \& environment & \makecell{number of people in homeless shelters,\\ percent of officers assigned to drug units, etc.,}                           \\ 
16                       & Income                     & \makecell{median household income, per capita income,\\ number of people under the poverty level, etc.,}                            \\ 
11                       & Education \& Employment                     & \makecell{percent of people 16 and over in the labor force and unemployed,\\ percent of people who do not speak English well, etc.,}                           \\ 
11                       & Family status                     & \makecell{percent of males who are divorced,\\ percent of moms of kids 6 and under in labor force, etc.,}                          \\ 
14                       & Immigrants                     & \makecell{total number of people known to be foreign-born,\\ percent of the population who have immigrated within the last 3 years, etc.,}                           \\ 
27                       & Housing                     & \makecell{percent of family households that are large (6 or more),\\ median gross rent as a percent of household income, etc.,}                            \\ 
12                       & Population                     & \makecell{population of the community, mean people per household, \\ number of people living in areas classified as urban, etc., }                            \\ \bottomrule
\end{tabular}
\end{table}

First, we remove missing values and features with duplicate information (e.g., retaining the percentage rather than the absolute number of populations below the poverty level) to reduce correlation. Next, logarithmic transformation is applied to features related to absolute income and population. Finally, we normalize each feature by $\Tilde{x}_{ij} = \frac{{x}_{ij} - \min\limits_i {x}_{ij}}{\max\limits_i {x}_{ij} - \min\limits_i {x}_{ij}}$ to ensure all feature values fall within the range $[0,1]$. After this pre-processing, a total of $n = 1993$ samples with $p = 97$ features, assumed to be distributed across $m = 7$ departments, are included in our analysis.

\begin{figure}[h]
  \centering
  \includegraphics[width=\textwidth]{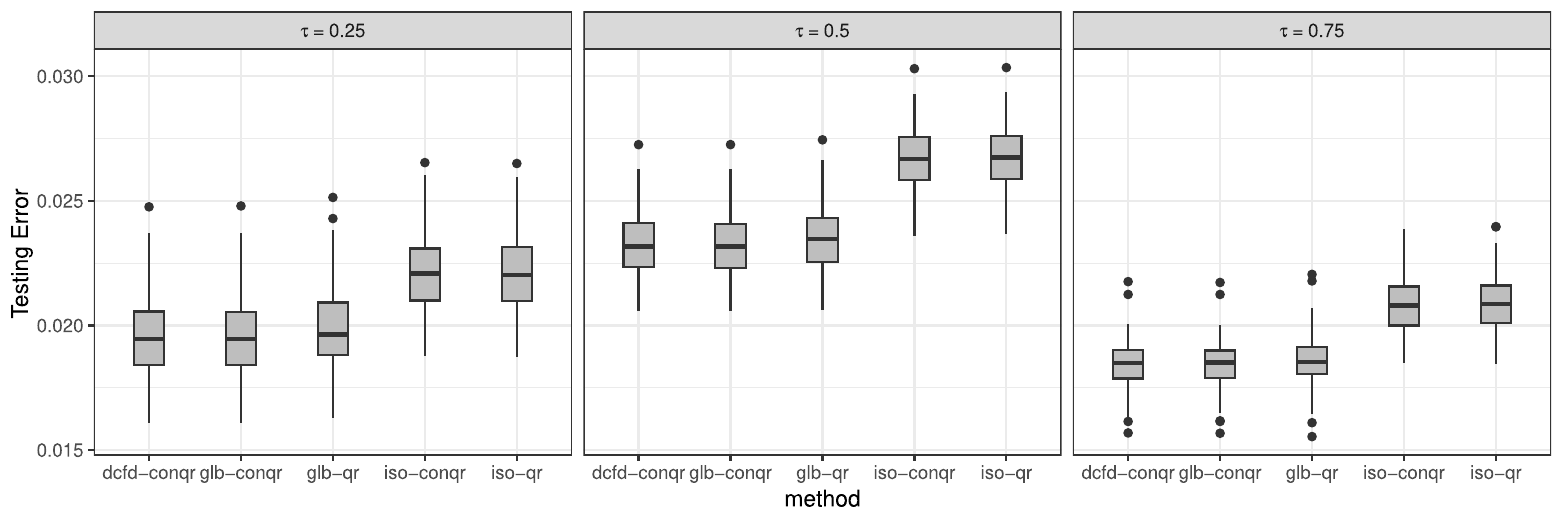}
  \caption{Box-plot of the testing error for the Communities and Crime dataset.}
  \label{Realdata_MTE}
\end{figure}

We assess estimation efficiency by randomly splitting the data into training ($\mathcal{S}_{\mathrm{train}}$) and testing ($\mathcal{S}_{\mathrm{test}}$) sets, with proportions of 90\% and 10\%, respectively. The estimation procedure is performed on the training set using \texttt{DSG-cqr} and the four competing methods, as described in the simulation section, with the isolated estimator calculated based on the first machine (the public facilities \& environment group). This random partition is repeated 100 times and the testing error is recorded as $\frac{1}{n_{\mathrm{test}}} \sum\limits_{i\in \mathcal{S}_{\mathrm{test}}} \rho_\tau(\hat{y}_{i} - y_{i})$, where $\hat{y}_{i}$ is calculated based on the estimator from $\mathcal{S}_{\mathrm{train}}$. Figure \ref{Realdata_MTE} presents box plots of the testing error for $\tau \in \{0.25, 0.5, 0.75\}$. The results indicate that the \texttt{DSG-cqr} estimator performs comparably to the global estimator and significantly outperforms the isolated estimator. Additionally, the mean testing error based on \texttt{conquer} is slightly lower than that of the ordinary quantile estimator across all cases, aligning with Table \ref{CVERR}. 

\begin{table}[h]\scriptsize
\setlength{\tabcolsep}{4pt}
\caption{95\% confidence intervals for the regression coefficients within the public facilities \& environment group. The full variable names are: \emph{NumInShelters} - number of people in homeless shelters; \emph{NumStreet} - number of homeless people counted in the street; \emph{LandArea} - land area in square miles; \emph{PopDens} - population density in persons per square mile; \emph{PctUsePubTrans} - percent of people using public transit for commuting; \emph{LemasPctOfficDrugUn} - percent of officers assigned to drug units.}
\label{Emp_CI}
\begin{tabular}{cccccccc}
\toprule
\multirow{2}{*}{$\tau$}  & \multirow{2}{*}{Method} & \multicolumn{6}{c}{Variable name (6 variables in the public facilities \& environment group)}                                                                                             \\
                      &                         & \emph{NumInShelters}      & \emph{NumStreet}          & \emph{LandArea}           & \emph{PopDens}            & \emph{PctUsePubTrans}      & \emph{LemasPctOfficDrugUn} \\ \midrule
\multirow{4}{*}{0.25} & \texttt{DSG-cqr(hr)}                       & {[}0.005,0.082{]}  & {[}0.017,0.085{]}  & {[}0.040,0.129{]}   & {[}0.056,0.068{]}  & {[}-0.101,-0.031{]} & {[}0.014,0.080{]}    \\
                      & \texttt{DSG-cqr(hs)}                       & {[}-0.004,0.091{]} & {[}-0.022,0.124{]} & {[}-0.007,0.177{]} & {[}0.056,0.068{]}  & {[}-0.101,-0.031{]} & {[}-0.004,0.098{]}  \\
                      & \texttt{glb-cqr}                     & {[}0.007,0.095{]}  & {[}0.021,0.080{]}   & {[}0.023,0.133{]}  & {[}-0.011,0.104{]} & {[}-0.123,-0.012{]} & {[}0.022,0.089{]}   \\
                      & \texttt{glb-qr}                       & {[}0.009,0.093{]}  & {[}0.027,0.074{]}  & {[}0.027,0.130{]}   & {[}-0.008,0.100{]}   & {[}-0.125,-0.010{]}  & {[}0.023,0.089{]}   \\
\multirow{4}{*}{0.5}  
                      & \texttt{DSG-cqr(hr)}                       & {[}-0.017,0.062{]} & {[}-0.020,0.116{]}  & {[}-0.014,0.133{]} & {[}0.015,0.028{]}  & {[}-0.094,-0.027{]} & {[}-0.013,0.068{]}  \\
                      & \texttt{DSG-cqr(hs)}                       & {[}-0.026,0.071{]} & {[}-0.026,0.123{]} & {[}-0.034,0.153{]} & {[}0.015,0.028{]}  & {[}-0.096,-0.025{]} & {[}-0.025,0.080{]}   \\
                      & \texttt{glb-cqr}                       & {[}-0.014,0.054{]} & {[}-0.026,0.087{]} & {[}0.001,0.161{]}  & {[}-0.040,0.061{]}  & {[}-0.092,0.002{]}  & {[}-0.011,0.060{]}   \\
                      & \texttt{glb-qr}                       & {[}-0.013,0.053{]} & {[}-0.015,0.077{]} & {[}-0.056,0.219{]} & {[}-0.039,0.061{]} & {[}-0.091,0.001{]}  & {[}-0.010,0.058{]}   \\
\multirow{4}{*}{0.75} & \texttt{DSG-cqr(hr)}                       & {[}-0.044,0.067{]} & {[}-0.020,0.190{]}   & {[}-0.097,0.184{]} & {[}-0.001,0.011{]} & {[}-0.096,-0.046{]} & {[}-0.042,0.056{]}  \\
                      & \texttt{DSG-cqr(hs)}                       & {[}-0.036,0.058{]} & {[}0.013,0.157{]}  & {[}-0.047,0.134{]} & {[}-0.001,0.011{]} & {[}-0.106,-0.037{]} & {[}-0.044,0.058{]}  \\
                      & \texttt{glb-cqr}                       & {[}-0.043,0.060{]}  & {[}-0.004,0.190{]}  & {[}-0.093,0.207{]} & {[}-0.047,0.070{]}  & {[}-0.109,-0.0140{]} & {[}-0.062,0.038{]}  \\
                      & \texttt{glb-qr}                       & {[}-0.037,0.054{]} & {[}-0.008,0.194{]} & {[}-0.047,0.161{]} & {[}-0.042,0.065{]} & {[}-0.107,-0.016{]} & {[}-0.044,0.021{]}  \\ \bottomrule
\end{tabular}
\end{table}

Next, we construct confidence intervals for the regression coefficients within the first group using our decentralized feature-distributed inference method, as well as the global method described in the simulation section. Before proceeding, we calculate the correlation between the variables in the first group and those in other groups; the mean (maximum) Pearson correlation coefficient is 0.117 (0.524). We then compute the confidence intervals for different quantile levels $\tau \in \{0.25, 0.5, 0.75\}$, and present the results in Table \ref{Emp_CI}. The findings indicate that our distributed inference method performs comparably to the global method, with the observed differences attributable to the mild correlation structure of the covariates. Additionally, the \texttt{conquer}-based methods yield wider confidence intervals than the ordinary quantile estimator, consistent with the discussion in \cite{Fernandes2021}.

\section{Conclusion}\label{sec:conclu}

This study introduces a novel decentralized surrogate gradient descent algorithm, \texttt{DSG-cqr}, for QR modeling in feature-distributed datasets under a decentralized network. The Gaussian mechanism is employed to ensure local differential privacy. We construct confidence intervals for the local parameters using the local auxiliary variables in \texttt{DSG-cqr}. Our results show that the proposed \texttt{DSG-cqr} achieves linear convergence to optimal statistical precision, outperforming estimators based on local datasets, particularly in the presence of potential model misspecification. Consequently, \texttt{DSG-cqr} enables statistically efficient joint QR modeling in feature-distributed datasets within a scalable, decentralized network. Since only auxiliary variables are transmitted between machines, the sensitive information in local raw data is effectively protected. Through finite sample simulations, we have demonstrated the influence of algorithmic parameters on finite sample performance. The proposed methodologies significantly outperform those based on local datasets and closely align with global estimators, consistent with the theoretical properties.

\bibliographystyle{elsarticle-harv} 
\bibliography{citation}

\end{document}